\documentclass[%
 pra,
 amsmath,amssymb,
 reprint,nofootinbib,%
]{revtex4-2}

\usepackage[utf8]{inputenc}

\usepackage{bm}
\usepackage{booktabs}
\usepackage{dcolumn}
\usepackage[T1]{fontenc}
\usepackage{graphicx}
\usepackage{hyperref}
\hypersetup{colorlinks=true,linkcolor=blue,citecolor=blue,filecolor=blue,urlcolor=blue}
\usepackage{listings}
\usepackage{mathptmx}
\usepackage{physics}
\usepackage{dsfont}
\usepackage{braket}
\usepackage{wasysym}
\usepackage{xcolor}
\definecolor{boxcolor}{RGB}{204, 222, 229}
\definecolor{framecolor}{RGB}{0, 91, 127}
\usepackage[most]{tcolorbox}

\lstdefinestyle{python}{
  language=Python,
  basicstyle=\ttfamily\footnotesize,
  commentstyle=\color{olive},
  keywordstyle=\color{blue},
  numberstyle=\tiny\color{gray},
  stepnumber=1,
  showstringspaces=false,
  tabsize=4,
  breaklines=true,
  columns=flexible,
}

\newcommand{\xdata}{\mathbf{x}}
\newcommand{\xdataset}{X}
\newcommand{\ydataset}{y}

\newtcolorbox[auto counter]{infobox}[3][]{%
    float,
    colback=boxcolor,
    colframe=framecolor,
    title={Info Box \thetcbcounter: #2},
    label={box:#3},
    #1
}

\begin{document}

\title{sQUlearn -- A Python Library for Quantum Machine Learning}

\author{David A. Kreplin}
\email{david.kreplin@ipa.fraunhofer.de}
\affiliation{Fraunhofer Institute for Manufacturing Engineering and Automation IPA, Nobelstraße 12, D-70569 Stuttgart, Germany}

\author{Moritz Willmann}
\thanks{These authors contributed equally}
\affiliation{Fraunhofer Institute for Manufacturing Engineering and Automation IPA, Nobelstraße 12, D-70569 Stuttgart, Germany}

\author{Jan Schnabel}
\thanks{These authors contributed equally}
\affiliation{Fraunhofer Institute for Manufacturing Engineering and Automation IPA, Nobelstraße 12, D-70569 Stuttgart, Germany}

\author{Frederic Rapp}
\thanks{These authors contributed equally}
\affiliation{Fraunhofer Institute for Manufacturing Engineering and Automation IPA, Nobelstraße 12, D-70569 Stuttgart, Germany}

\author{Manuel Hagelüken}
\thanks{These authors contributed equally}
\affiliation{Fraunhofer Institute for Manufacturing Engineering and Automation IPA, Nobelstraße 12, D-70569 Stuttgart, Germany}

\author{Marco Roth}
\email{marco.roth@ipa.fraunhofer.de}
\affiliation{Fraunhofer Institute for Manufacturing Engineering and Automation IPA, Nobelstraße 12, D-70569 Stuttgart, Germany}

\date{\today}

\begin{abstract}
sQUlearn introduces a user-friendly, NISQ-ready Python library for quantum machine learning (QML), designed for seamless integration with classical machine learning tools like scikit-learn. The library's dual-layer architecture serves both QML researchers and practitioners, enabling efficient prototyping, experimentation, and pipelining. sQUlearn provides a comprehensive toolset that includes both quantum kernel methods and quantum neural networks, along with features like customizable data encoding strategies, automated execution handling, and specialized kernel regularization techniques. By focusing on NISQ-compatibility and end-to-end automation, sQUlearn aims to bridge the gap between current quantum computing capabilities and practical machine learning applications. The library provides substantial flexibility, enabling quick transitions between the underlying quantum frameworks Qiskit and PennyLane, as well as between simulation and running on actual hardware.
\end{abstract}

\maketitle

\section{Introduction}

Machine Learning (ML) is a remarkably successful discipline that has been rapidly adopted broadly in science, industry and society, and is widely believed to have the potential to completely transform a wide range of industries in the upcoming years~\cite{chui2023}. While its recent ascent to prominence is notable, the roots of ML extend back to the early 1960s~\cite{Samuel1956}, reflecting a path of varied progress and challenges. Besides some initial set-backs~\cite{Crevier1993} the foundations of today's deep neural networks that drive a lot of the recent advancements have already been established almost half a century ago~\cite{Linnainmaa1976, Werbos1982}. The reasons for the accelerated breakthroughs of the past decade are essentially threefold: \hypertarget{ref:compute}{(a)} an increased computational resources, \hypertarget{ref:data}{(b)} the availability of large-scale data, and \hypertarget{ref:tools}{(c)} the emergence of development tools that abstract away low-level complexity. While these factors have enabled remarkable breakthroughs, the evolving landscape of ML presents new frontiers, one of which is quantum machine learning (QML).

Quantum machine learning has emerged as an innovative approach that explores different capabilities and potentials within the field, leveraging the principles of quantum mechanics to enhance computational power and efficiency~\cite{Cerezo2022}. Some techniques seek to accelerate performance by executing quantum variants of linear algebra procedures~\cite{Harrow2009, Rebentrost2014, Zhao2019, Liu2023}, effectively using the quantum computer as an hardware accelerator, similar to a GPU. However, these methods usually involve deep quantum circuits with many gates. The resulting complexity often exceeds the capabilities of current noisy intermediate-scale quantum (NISQ) hardware~\cite{Preskill2018}. This has spawned an increased interest in NISQ-compatible models which are often not just quantum-enhanced versions of classical algorithms but rather models that have an intrinsic quantum nature. Quantum kernel methods, for example, have been shown to have the potential to outperform their classical counterparts in specific tasks, making QML an attractive domain for further research and practical applications~\cite{Liu2021, Huang2022}.

Despite these successes, the situation of QML today is comparable to that of classical ML a few decades ago. Translating the success factors \hyperlink{ref:compute}{(a)}--\hyperlink{ref:tools}{(c)} of classical ML to QML, we observe that the progress of the hardware developments poses a major bottleneck for the adaptability of QML algorithms for practical usage [cf.~\hyperlink{ref:compute}{(a)}]. In terms of data [cf.~\hyperlink{ref:data}{(b)}], we find that  QML can mostly benefit from the tremendous amounts of classical data available, although this is not necessarily true for quantum data~\cite{Cerezo2022}. As for development tools, most of the tools for QML are extensions of low-level quantum computing packages that require exhaustive knowledge in quantum computing and ML and often require manipulations on the qubit level~\cite{QiskitCommunity2017, broughton2021tensorflow, bergholm2022pennylane, cirq2023, DiMarcantonio2023}. In this work, we aim to bridge the gap for the success factor~\hyperlink{ref:tools}{(c)} and introduce sQUlearn, an easy-to-use and NISQ-ready python library for QML which aims to democratizing access to QML.

Given the close relationship between the fields of ML and QML, we strive for high compatibility with already available tools. scikit-learn~\cite{Pedregosa2011} is a widely-used Python library for ML, well known for its simple and effective API~\cite{buitinck2013}. Leveraging this, sQUlearn provides an scikit-learn interface for QML methods which allows for a seamless integration into a wide range of available tools ranging from scikit-learn-based pipelines over MLOps~\cite{Kreuzberger2022, Zaharia2018} to AutoML~\cite{Zoeller2021, liaw2018tune}. Besides technical streamlining, this approach also has the benefit of a shallow learning curve for practitioners already familiar with tools for classical ML. The library offers a variety of high-level implementations of quantum neural networks (QNN)~\cite{Cerezo2021},  quantum convolutional neural networks (QCNN)~\cite{Cong_2019}, and quantum kernel methods~\cite{Schuld2019} in various flavors such as quantum support vector machines (QSVM)~\cite{Havlicek2019} and quantum Gaussian processes~\cite{Rapp2023} which can be utilized for classification and regression.

\begin{figure}
    \includegraphics[width=\columnwidth]{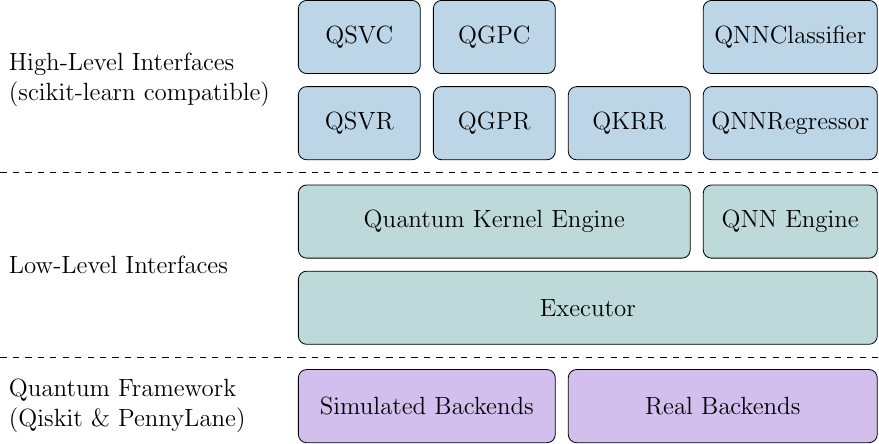}
    \caption{Schematic overview of sQUlearn. sQUlearn offers scikit-learn compatible high-level interfaces for various kernel methods and QNNs. They build on top of the low-level interfaces of the QNN engine (Sec.~\ref{sec:qnn_engine}) and the quantum kernel engine (Sec.~\ref{sec:kernel_engine}). The executor is used to run experiments on simulated and real backends in the Qiskit or PennyLane environment.}
    \label{fig:squlearn_architecture}
\end{figure}

These high-level methods are based on low-level implementations which provide advanced tools for QML researchers. These features include a flexible QNN engine that allows for arbitrary differentiation, tools for designing quantum circuits such as simple encoding circuit creation and analysis including the detection and removal of redundant parameters~\cite{Haug2021}.  In terms of quantum kernel methods both, fidelty-type (FQK)~\cite{Havlicek2019} and projected quantum kernels (PQK)~\cite{Huang2021} are featured in a general and versatile implementation. The execution on simulated or real quantum backends such as IBM Quantum\footnote{https://quantum.ibm.com/} or Amazon Braket\footnote{https://aws.amazon.com/braket/} is possible through the quantum frameworks Qiskit~\cite{QiskitCommunity2017} and PennyLane~\cite{bergholm2022pennylane}.
The centralized execution in sQUlearn allows for a quick transition from simulations to different real quantum computers, and include several convenience features such as automated session handling for Qiskit Runtime, storage of results or automatic restarts of failed jobs. 
An overview of sQUlearns structure is shown in Fig.~\ref{fig:squlearn_architecture}.

sQUlearn is developed to accommodate users with knowledge from diverse backgrounds, enabling effective utilization regardless of the user's primary field of expertise. For people already familiar with ML and scikit-learn, the manuscript provides colored info boxes which give a quick overview of how the high-level methods can be used. The remainder of this manuscript is structured as follows. Section~\ref{sec:similar_work} relates sQUlearn to similar work and frameworks. Section~\ref{sec:squlearn_library} introduces the fundamental concepts behind QNNs and quantum kernel methods and details their respective implementation in sQUlearn. An illustrative example involving some of sQUlearn's unique aspects is demonstrated in Sec.~\ref{sec:example}. Finally, Sec.~\ref{sec:conclusion} concludes the manuscript.

\section{Similar Work}
\label{sec:similar_work}

In the following, we provide an overview of the most commonly used QML-related Python packages. Afterwards, we highlight the unique features that set sQUlearn apart.

Qiskit~\cite{QiskitCommunity2017} is a general open-source software development kit for gate based quantum computing. It offers a direct hardware access to a wide selection of backends with superconducting qubits and several simulators. Qiskit has a dedicated ML package called "Qiskit Machine Learning" that offers various low-level implementations for quantum kernels and also a high-level QSVM implementation. The quantum kernels are restricted to fidelity kernels. Additionally, high- and low-level implementations of QNNs are available. Although the high-level implementation provides a simple interface, it enforces a strict separation of input encodings and the ansatz with trainable weights, limiting features such as combining input data with weights or using repeated layers for data re-uploading. While the low-level QNN implementation offers a large flexibility in the circuit design, derivatives are, however, limited to first-order due to the absence of higher-order derivatives in Qiskit primitives. Furthermore, the observables in the QNN can not be parameterized, making them non-trainable.

TensorFlow Quantum~\cite{broughton2021tensorflow} is an extension of Cirq~\cite{cirq2023} that primarily focusses on low-level development. It integrates with the classical ML library TensorFlow, enabling automatic gradient back-propagation for quantum circuits~\cite{tensorflow2015}. This facilitates the training of hybrid quantum-classical ML models and allows the use of the Keras API for QML models~\cite{Chollet2015}. TensorFlow Quantum focuses on QNNs and does not provide pre-built quantum kernel methods. Cirq, on the other hand, is a general quantum computing framework for gate-based quantum computing, offering the capability to implement quantum circuits and run them on various simulators and hardware architectures. 

PennyLane~\cite{bergholm2022pennylane} is a quantum computing framework with a focus on differentiable programming and QML. It offers automatic differentiation for quantum circuits, enabling the implementation of quantum circuits for various hybrid QML methods. PennyLane furthermore offers multiple templates for data encoding and entangling layers, that can be used to construct a parameterized quantum circuit. It also offers integration into classical ML libraries such as PyTorch~\cite{Paszke2019} and Keras allowing the construction of complex and hybrid ML models. However, simple high-level interfaces are not available. Nevertheless, quantum kernels constructed using PennyLane can be utilized as input for scikit-learn's high-level kernel implementation. PennyLane offers various built-in simulators as well as a plugin structure to connect to other simulator- and hardware architectures.
The usage of automatic differentiation libraries enables a fast computation of derivatives of quantum circuits for noise-free simulations.

QuASK (Quantum Advantage Seeker with Kernels)~\cite{DiMarcantonio2023} is a python framework to run QML experiments with quantum kernels. The software is built on top of PennyLane for defining quantum circuit software and utilizes scikit-learn. It is designed to automate each phase of a quantum kernel QML experiment from the selection of the dataset (both classical and quantum) and its preprocessing to calculating metrics of resulting Gram matrices such as the geometric difference~\cite{Huang2021} to provide an assessment about potential quantum advantage. Parameterized quantum kernels (FQK and PQK) can be trained using gradient-descent-based optimization, grid search or genetic algorithms. The PQK implementation of QuASK is restricted to linear PQKs and radial basis function (RBF) outer kernels, both based on one-particle reduced density matrices (see Sec.~\ref{subsubsec:kernel-background} for details). 

sQUlearn distinguishes itself from these frameworks in several aspects:
It is the only framework that provides a unified high- and low-level implementation for fidelity and projected quantum kernels as well as QNNs, all implemented with Qiskit and PennyLane.
By leveraging both frameworks, sQUlearn not only grants convenient access to hardware via Qiskit but also enables rapid derivative computation with simulators in PennyLane.
A seamless transition between simulation and hardware access is enabled by a centralized execution engine. 
Furthermore, it offers several unique features such as the simplified generation of encoding circuits from strings, flexible QNN architectures capable of calculating arbitrary derivatives, automatic shot-adjustments and variance-regularization. 
Additionally, sQUlearn provides an extensive flexibility in creating PQKs. A large variety of outer kernels beside the RBF kernel are available, as for example the Matern kernel or the rational quadratic kernel.
Moreover, with its automated session handling for real quantum hardware, sQUlearn has a focus on employing and training ML models on currently available backends, allowing users to explore NISQ hardware while focusing on problem-solving rather than managing computational resources. sQUlearn serves as a one-stop-shop for today's QML algorithms, integrating them into a single library and offering a seamless interface with classic ML packages like scikit-learn. Finally, through its dual-layer approach sQUlearn caters to both QML researchers and practitioners. Advanced users can access low-level functionalities for research, while practitioners can quickly implement ML models using high-level APIs and the wide variety of available models.

\section{The sQUlearn Library}
\label{sec:squlearn_library}

In the following, we consider supervised learning problems with a classical\footnote{Although we assume classical data here, sQUlearn can cope with quantum states as input as well.} dataset $\mathcal{D}=(\xdataset,\ydataset)$, where $\xdataset=\lbrace \xdata_1,\dots,\xdata_N\rbrace$ is a set of $N$ data points $\xdata_i\in\mathcal{X}$ with labels $y_i\in\mathcal{Y}$. Here, $\mathcal{X}$ is the data space which is here chosen as a subspace of $\mathbb{R}^d$ for simplicity. Typical scenarios for the output domain include regression problems with $\mathcal{Y}\subset{\mathbb{R}}$ and classification tasks with $\mathcal{Y}\subset\mathbb{N}$, although sQUlearn supports multidimensional output as well. 

To enable quantum processing, the data is encoded into a quantum state
\begin{equation}
    \ket{\psi(\xdata,\pmb{\theta})}=U(\xdata, \pmb{\theta})\ket{0}\,,
\label{eq:encoding_circuit}
\end{equation}
In practise the encoding operator $U(\xdata,\pmb{\theta})$ is implemented by a quantum circuit which manipulates an initial $n$-qubit quantum state $\ket{0}$ based on the data $\xdata\in\mathcal{X}$. While various proposals for encoding such as encoding data directly into the amplitudes of the wavefunction or bit-wise into the basis states have been suggested~\cite{Schuld2018}, the predominant approach for NISQ compatible algorithms is based on encoding the components $x^i$ of the data vectors $\xdata$ as angles in one or two-qubit rotations. Typically, these operations are supplemented with data independent gates parameterized by some parameters $\pmb{\theta}=(\theta_1,\dots,\theta_K)$, with $\theta_i\in\mathbb{R}$. The resulting circuit is often referred to as encoding circuit. Since this is the defining object for a wide range of algorithms, sQUlearn provides several pre-implemented encoding circuits (see Info Box~\ref{box:encoding_circuits}) and offers a variety of tools for analysing and creating custom encodings.
Redundant parameters can be identified and removed automatically from the encoding circuit. This is achieved by utilizing the quantum Fisher information matrix \cite{Liu2020,Meyer2021} and an black-box implementation of the algorithm introduced in Ref. \onlinecite{Haug2021}.
Furthermore, differentiation with respect to arbitrary parameters or data vector components is implemented for encoding circuits by employing the parameter-shift rule.

\begin{infobox}{Data Encoding and Feature Maps}{encoding_circuits}
Data encoding circuits are used to embed classical data into a quantum state and manipulate the quantum state via trainable parameters $\pmb{\theta}$.
They are a key component of many QML algorithms, and the design of a good data encoding circuit is crucial for the performance of the algorithm.
In sQUlearn, encoding circuits are a mandatory input for QNNs or quantum kernel programs.
sQUlearn offers a wide range of pre-implemented quantum encoding circuits, which can be combined to create more sophisticated encodings.
For example, the encoding circuit introduced in Ref. \onlinecite{Peters2021} can be created and displayed with the following code 
\begin{lstlisting}[style=python]
from squlearn.encoding_circuit import (
    HighDimEncodingCircuit)

encodig_circuit = HighDimEncodingCircuit(
    num_qubits=5, num_features=23, num_layers=2)
encodig_circuit.draw("mpl")
\end{lstlisting}
\includegraphics[width=\linewidth]{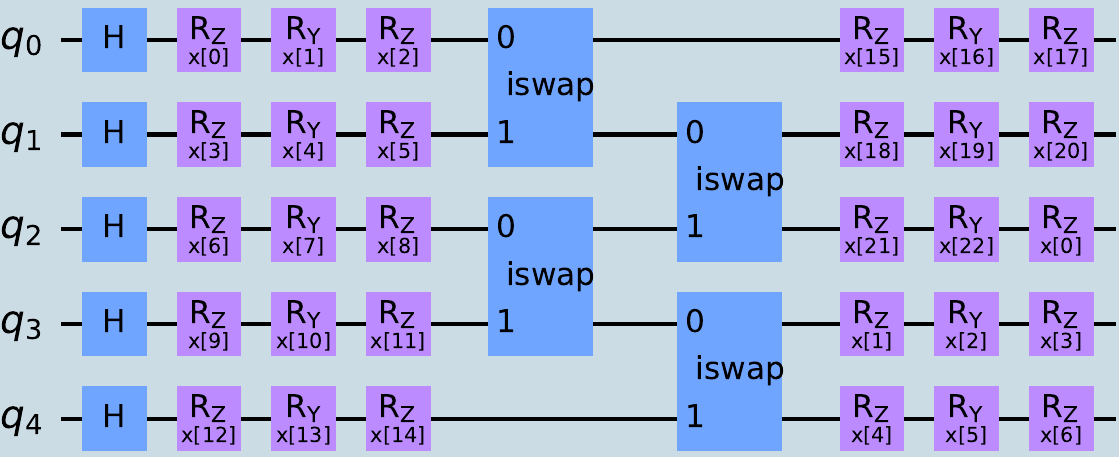}
\end{infobox}

The two core classes of algorithms in sQUlearn are QNNs and quantum kernel methods. While both methods can be shown to be equivalent in theory~\cite{schuld2021supervised}, in practise they have their individual advantages and disadvantages. In the first approach, short-depth parametrized quantum circuits are executed on a quantum computer, with the optimization of the parameters being conducted classically. The second approach relies on the quantum computer only for evaluating the Gram matrix, while the subsequent processes typically employ established classical algorithms such as support vector machines (SVM)~\cite{vapnik1999} or kernel ridge regression (KRR)~\cite{Murphy2012}. In sQUlearn, QNNs and kernel methods are driven by two separate low-level implementations which will be explained in the following.

\subsection{Quantum Neural Network Engine}
\label{sec:qnn_engine}
\begin{figure*}
	\includegraphics[width=0.8\linewidth]{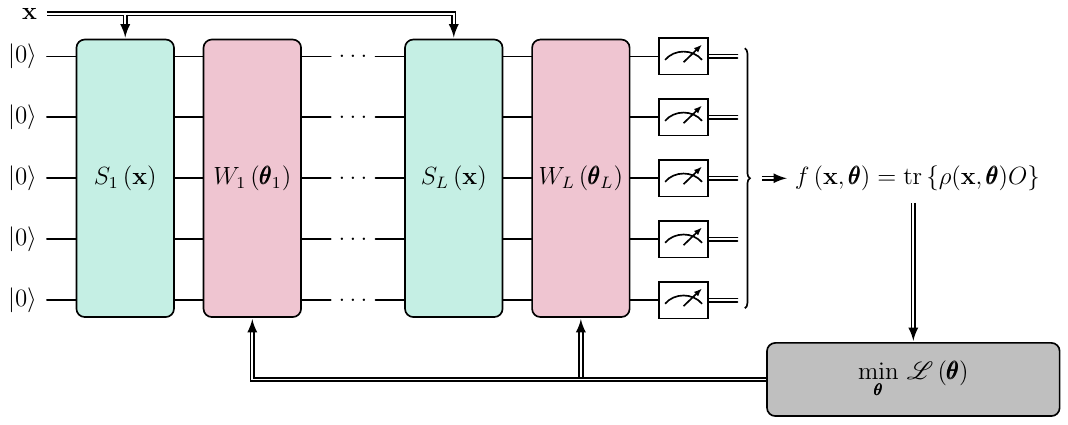}
	\caption{Prototypical architecture of a QNN. Classical data is embedded into quantum states using encoding operations $S_l(\xdata)$. Parameterized quantum operations $W_l(\theta)$ are then applied to process the data. Often, this structure is repeated for several layers. After the final layer, a measurement is performed and an expectation value $f$ is calculated, see Eq.~\eqref{eq:quantum-model-general}. The parameters $\pmb{\theta}$ are then optimized using a classical optimizer to minimize a suitable empirical loss function $\mathcal{L}$, Eq.~\eqref{eq:empirical_loss}. The figure is adapted freely from Ref.~\cite{Mangini2021}.
 }
	\label{fig:qnn_structure}
\end{figure*}
In this section, we briefly introduce the basic principles of QNNs and demonstrate how they are implemented in the sQUlearn library.
\subsubsection{Background}
Quantum neural networks are variational models that calculate an expectation value to optimizes a cost function. An example of a QNN can be seen in Fig.~\ref{fig:qnn_structure}. Data is embedded into a quantum state using a data encoding circuit Eq.~\eqref{eq:encoding_circuit}. The encoding circuit $U(\xdata,\pmb{\theta})$ often comprises a layered structure $U(\xdata,\pmb{\theta})=\prod_{l=1}^L S_l(\xdata)W_l(\pmb{\theta}_l)$, where $S_i(\xdata)$ denotes a data-dependent operation,\footnote{In principle, $S_j(\xdata)$ can depend on additional parameters $S_j(\xdata)\equiv S_j(\xdata,\vartheta)$.} $W_i(\pmb{\theta}_i)$ are parameterized operators and $L$ is the number of layers. The resulting state $\rho(\xdata,\pmb{\theta})$ is measured repeatably to calculate the expectation value of some observable $O$
\begin{equation}
    \label{eq:quantum-model-general}
    f(\xdata,\pmb{\theta}) = \tr\left(\rho(\xdata,\pmb{\theta})O\right)\,.
\end{equation}
The parameters $\pmb{\theta}$ are optimized by minimizing the empirical loss
\begin{equation}
    \label{eq:empirical_loss}
    \mathcal{L}(\pmb{\theta})=\frac{1}{N}\sum_{i=1}^N \ell(\xdata_i, y_i, \pmb{\theta})\,,
\end{equation}
where $\ell$ is a suitable loss function such as the Euclidean distance between $f(\xdata_i)$ and $y_i$. The minimization of $\mathcal{L}$ is typically performed utilizing gradient-based techniques like stochastic gradient descent \cite{Harrow2021}. The gradient is usually obtained using either finite differences or by exploiting the parameter-shift rule~\cite{Mitarai2018}. If a parameter $\theta$ appears in a single rotation gate $R_\alpha(\theta) = \exp(-i \theta \sigma_\alpha/2)$, with $\sigma_\alpha$ being a Pauli operator with $\alpha=x,y,z$, the parameter-shift rule can be expressed as follows~\cite{Mitarai2018,Kyriienko2021}:
\begin{equation}
\frac{d}{d\theta} \braket{\tilde{\psi}|{R}_\alpha[\varphi(\theta)]{\tilde{O}}{R}_\alpha[\varphi(\theta)]|\tilde{\psi}} \!=\! \frac{1}{2} \varphi'(\theta)\!\big[\! \braket{{O}}^+ \!\!\!- \braket{{O}}^- \!\big]\!.
\end{equation}
Here, the term $\braket{{O}}^\pm$ denotes the expectation value of ${O}$ with the parameter $\theta$ shifted by $\pm \pi/2$:
\begin{equation}
\braket{{O}}^\pm = \braket{\tilde{\psi}|{R}_\alpha(\varphi(\theta)\pm \tfrac{\pi}{2}){\tilde{O}}{R}_\alpha(\varphi(\theta)\pm \tfrac{\pi}{2})|\tilde{\psi}}.
\end{equation}
In this example, the function $\varphi(\theta)$ is an optional and typically non-linear map for the input data.
The operator ${\tilde{O}}$ encompasses all gates applied after the rotation gate ${R}_\alpha$ that depends on $\theta$. Similarly, $\ket{\tilde{\psi}}$ includes all gates applied before the rotation gate. 
If the parameter $\theta$ is present in multiple gates, the product rule of differentiation is applied, resulting in an additional summation over all occurrences. Higher-order derivatives can be obtained through successive applications of the parameter-shift rule. The parameter-shift rule yields an exact gradient and calls for the same two circuit evaluations as the finite differences approach when the parameters are exclusively found in a single rotation gate.

Quantum neural networks have been applied to a variety of problems~\cite{Cerezo2022,Gujju2023} and have been shown to have interesting properties in terms of expressivity~\cite{Schuld2021} and data efficiency~\cite{Caro2022}. 
sQUlearn offers a custom low-level QNN implementation based on Qiskit circuits and observables, and with support for the fast automatic differentiation in simulated backends of PennyLane. 
One of the core features of sQUlearn's QNN engine is the ability to take arbitrary derivatives which readily allows for the application in advanced tasks such as solving differential equations~\cite{Kyriienko2021}. The training process is significantly streamlined and automated, allowing for the use of various classical optimizers, mini-batching and stochastic gradient descent and advanced regularization techniques such as the one in~\cite{Kreplin2023}. The QNN implementation allows for a straightforward generalization to multiple outputs~\cite{squlearnDocu2023}. In designing sQUlearn, we have paid particular attention to facilitate training on real quantum hardware, as detailed in Sec.~\ref{sec:executor}.

Furthermore, sQUlean provides an implementation for QCNNs as an encoding circuit.
QCNNs have been shown to be efficiently trainable, as the number of parameters increases logarithmically with the number of qubits~\cite{Pesah2021}.
They are inspired from their classical counterparts and feature a sequence of convolution and pooling layers that are available as building blocks in the QCNN encoding circuit class, simplifying the construction.  
Pooling layers in QCNNs reduce the qubit count, leading to the dimensionality reduction as implemented in classical convolutional neural network. This reduction can be accomplished either through measurement that collapse the quantum state and allowing for conditional gate operations based on the measurement results, or by employing entangling gates that consolidate the information from multiple qubits into a smaller subset. Both variants are readily implemented in sQUlearn.
A simple classification example utilizing an QCNN encoding circuit is shown in Appendix~\ref{sec:appendix_qcnn_example}.

\begin{infobox}{QNNs and Observables}{qnn}

In sQUlearn, a QNN is created when observables are combined with a trainable encoding circuit. Observables are quantum mechanical operators employed to determine expectation values of a quantum state. In sQUlearn, observables can be constructed from Pauli operators. A variety of preconfigured observables is available, and in addition, custom observables can be effortlessly constructed from strings. Furthermore, arithmetic operations between observables are also implemented.
Training the QNN optimizes the trainable parameters of the encoding circuit and the observables to minimize a loss function that quantifies the quality of the QNN.
Derivatives with respect to parameters or input data can be obtained using the parameter-shift rule. The construction and training of a QNN in sQulearn can be achieved with just a few lines of code:
\begin{lstlisting}[style=python]
from squlearn import Executor
from squlearn.encoding_circuit import (
    YZ_CX_EncodingCircuit)
from squlearn.observables import IsingHamiltonian
from squlearn.optimizers import Adam
from squlearn.qnn import QNNRegressor, SquaredLoss

pqc = YZ_CX_EncodingCircuit(
    num_qubits=5, num_features=2, num_layers=2)
obs = IsingHamiltonian(num_qubits=5)
qnn = QNNRegressor(
    pqc, obs, Executor(), SquaredLoss(), Adam())
qnn.fit(x_train, y_train)
y_test = qnn.predict(x_test)
\end{lstlisting}
The example first initializes the encoding circuit introduced in Ref. \onlinecite{Haug2023} and an Ising Hamiltonian as an observable. Afterwards the \texttt{QNNRegressor} is initialized; alternatively the classifier
\texttt{QNNClassifier} could be used for a classification problem. 
The training of the QNN with training data \texttt{x\_train} and labels \texttt{y\_train} is executed by calling the \texttt{fit} function.
The last line shows the inference of the QNN on some test data \texttt{x\_test} utilizing the \texttt{predict} method.   
\end{infobox}

\subsubsection{Implementation \label{sec:qnn_implementation}}

In sQUlearn, QNNs are constructed from parameterized data encoding circuits $U(\xdata, \pmb{\theta})$ and observables $O$ for defining the outputs, see Eqs.~\eqref{eq:encoding_circuit} and \eqref{eq:quantum-model-general}.
sQUlearn provides dedicated high-level data structures that enable the convenient utilization and construction of these elements, leveraging Qiskit's circuits and operators underneath.
Complex operators or encoding circuits can be created by assembling predefined components or by employing custom implementations specified through strings.
The conversion to PennyLane's circuit and operator format is implemented, allowing to take advantage its fast automatic differentiation in simulations. The conversion also supports non-linearly parameterized gates.
Additionally, sQUlearn offers flexibility in adjusting hyperparameters after initializing the QNN, such as the number of qubits or the number of layers in the encoding circuit.
It is possible to optimize these hyperparameters through methods like grid search implemented in scikit-learn~\cite{Pedregosa2011} or similar approaches.
This is demonstrated for PQKs (which are based on the low-level QNN engine) in Sec.~\ref{sec:example}.

\begin{figure}
\includegraphics[width=0.9\linewidth]{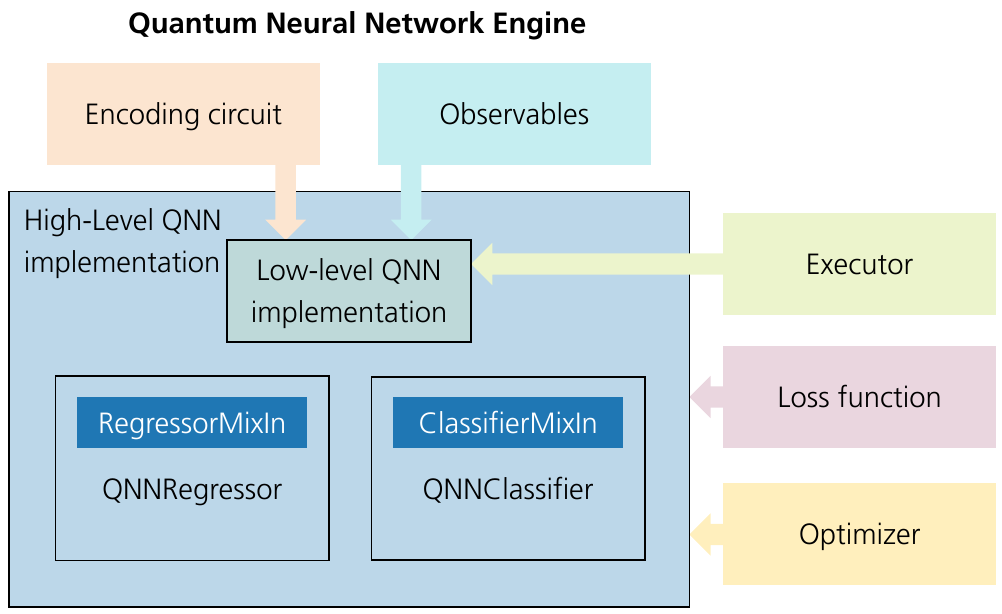}
\caption{
Schematic overview of the QNN engine in sQUlearn:
The central box in the diagram represents the high-level implementation that features the training and evaluation of QNNs with the scikit-learn interface.
The elements surrounding the box are essential for initializing the high-level methods, and include the encoding circuit, output observables, the Executor, the loss function, and the optimizer.
Within the high-level implementation, the encoding circuit, observables, and the Executor are utilized to set-up the low-level QNN implementation which leverages either PennyLane or Qiskit depending on the chosen backend of the Executor. This low-level implementation is also available as a standalone object in sQUlearn, and it is responsible for the evaluation of the QNN and its derivatives.
Utilizing the scikit-learn RegressorMixin or ClassifierMixin creates the sQUlearn high-level QNN implementations, namely \texttt{QNNRegressor} and \texttt{QNNClassifier}, designed for regression and classification tasks.
\label{fig:highlevel_qnn}}
\label{fig:qnn_design}
\end{figure}

sQUlearn offers the computation of arbitrary derivatives with respect to both parameters and features of the data encoding circuits. Similarly, observables can be parameterized and are fully differentiable as well.
When utilizing a PennyLane device in sQUlearn, the derivatives are computed by PennyLane's built-in differentiation routines.
In the case of the Qiskit implementation, a custom data structure has been developed to compute arithmetic operations of circuit outputs and expectation values for realizing arbitrary derivatives. 
This implementation is tailored to evaluate and train QNNs on real quantum hardware more efficiently. 
We employ the parameter-shift rule to calculate circuit derivatives and apply it successively for higher-order derivatives.
Furthermore, we optimize efficiency of the evaluations by caching all circuit and observable derivatives, ensuring they are computed only once.

The evaluation of expectation values is facilitated using Qiskit's Estimator and Sampler primitives, all managed by our \texttt{Executor} class (refer to Sec.~\ref{sec:executor}).
The evaluation routines are optimized to ensure that each circuit is assessed only once, especially when measurements can be utilized in multiple expectation values.
This includes scenarios where there are multiple observables with the same measurement basis, observable derivatives, or variance calculations.
This feature is available when measuring probabilities utilizing the Sampler primitive instead of expectation values. It becomes particularly useful on real quantum hardware to reduce the number of costly circuit evaluations.
Furthermore, to minimize the need for repeated evaluations, all calculated expectation values are cached.
Error mitigation is provided through the Qiskit primitives, making all options implemented in Qiskit accessible, but for QNNs, it is often not needed in our experience. 
Instead, if the aim is to execute the QNN on a real quantum backend, it is highly advisable to train the QNN on that specific backend to enhance result accuracy. 
This approach allows for the optimization of parameters to effectively counteract systematic errors inherent in the real quantum hardware~\cite{McClean2016,Cerezo2021,Kreplin2023}. 

sQUlearn also offers two high-level implementations for QNNs and their training: the \texttt{QNNRegressor}, designed for optimizing regression tasks, and the \texttt{QNNClassifier}, tailored for classification problems, respectively (see Info box~\ref{box:qnn}).
To utilize these high-level routines, they need to be initialized by specifying an encoding circuit, observables, the Executor, the loss function, and an optimizer, as illustrated in Fig.~\ref{fig:highlevel_qnn}. Training and inference processes align with the familiar scikit-learn interface, using the canonical functions \texttt{fit} and \texttt{predict}.

In the following, we provide a code example illustrating how to set up and train a QNN.
In this specific setup, we solve a regression task for a one dimensional function.
The Executor class is initialized with 5000 shots and utilizes per default the PennyLane implementation for simulated results. 
We also incorporate variance regularization to mitigate the finite-sampling noise in the output and adaptively control the number of shots that are employed in the gradient circuit evaluation as outlined in Ref. \cite{Kreplin2023}. The full executable code is displayed in Appendix~\ref{section:appendix_qnn_example}. 

\begin{lstlisting}[style=python]
encoding_circuit = ChebyshevPQC(
    num_qubits=4, num_features=1, num_layers=3)

observable = SummedPaulis(num_qubits=4)

qnn = QNNRegressor(
    encoding_circuit,
    observable,
    Executor(shots=5000),
    SquaredLoss(),
    Adam(options={"lr": 0.3}),
    variance=0.005,
    shot_control=ShotsFromRSTD())

qnn.fit(x_train, y_train)
qnn.predict(x_test)
\end{lstlisting}

The example begins by initializing the encoding circuit with ten qubits and three layers.
The number of features is equal to one since $X$ is one dimensional. In this example, the same feature is encoded in gates applied to each of the ten qubits, repetitively. Next, the observable is defined as the sum of single Pauli operators, each acting on a single qubit (with the default being the Pauli-$Z$ operator).
The QNN for regression is set up using the \texttt{QNNRegressor} class. This includes the Executor, the loss function (squared error loss), and the optimizer (Adam).
Optional settings for variance regularization and adaptive shot control, named \texttt{variance} and \texttt{shot\_control}, are specified.
Finally, the QNN is trained on the training data (\texttt{x\_train} and labels \texttt{y\_train}) using the \texttt{fit} function. Predictions for a given test dataset (\texttt{x\_test}) are obtained with the \texttt{predict} method. 
An example for the \texttt{QNNClassifier} class and utilizing an QCNN encoding circuit is shown in Appendix~\ref{sec:appendix_qcnn_example}.

\subsection{Quantum Kernel Engine}
\label{sec:kernel_engine}
Quantum kernel methods are appealing because they allow QML to be formally embedded into the rich and powerful mathematical framework of classical kernel theory~\cite{schuld2021supervised, schölkopf2002learning}. The following first outlines the most important theoretical aspects of quantum kernels before discussing their implementation within the sQUlearn package as well as the interface to the respective classical kernel methods (cf. Info box~\ref{box:kernels}). 
\subsubsection{\label{subsubsec:kernel-background}Background}
The key idea of kernel methods is to find and analyze patterns by transforming data into a high (potentially infinite) dimensional feature space $\mathcal{F}$, where the learning problem attains a trivial form. The mapping of the data from the original input space $\mathcal{X}$ to $\mathcal{F}$ is accomplished by a feature map $\phi: \mathcal{X} \to \mathcal{F}$. The  feature space can be accessed through inner products of the feature vectors $\phi(\xdata)$, which is a function $K$ of two data points $\xdata, \xdata'$ called the kernel, i.e., 
\begin{equation}
    \label{eq:kernel}
    K(\xdata, \xdata') = \langle\phi(\xdata), \phi(\xdata')\rangle_{\mathcal{F}}\,.
\end{equation}
This operation avoids evaluating the feature map explicitly and thus makes the computation efficient. This is known as the kernel trick. One of the main achievements of classical kernel theory is the representer theorem~\cite{schölkopf2002learning}. It guarantees that the function which minimizes a regularized empirical risk Eq.~\eqref{eq:empirical_loss} can always represented as a finite (length of the training sample) weighted linear combination of the kernel between some $\xdata\in\mathcal{X}$ and the training data $X$. It can be shown that for a convex loss function, this is a convex optimization problem, which makes them interesting in principle, although they usually scale unfavorable with the amount of training data~\cite{schölkopf2002learning}.

Quantum kernel methods can be formulated as a classical kernel method (e.g. SVM) whose kernels are computed by a quantum computer. Using quantum kernels instead of classical kernels holds out the prospect of designing ML models that are able to deal with complex problems that are out of reach for conventional ML methods~\cite{Liu2021}. This is achieved by embedding data into a quantum state using a quantum feature map. In practice, this corresponds to applying a data encoding circuit as defined in Eq.~\eqref{eq:encoding_circuit}.
Note that here we have omitted the explicit dependence of the quantum state on (variational) parameters for clarity. 
Nevertheless, it is important to keep in mind that if the encoding circuit includes variational parameters $\pmb{\theta}$, the kernel matrix depends on these parameters, too. This can be exploited to create a more suitable embedding into the feature space, as shown in the example in Sec. \ref{subsubsec:kernel-implementation}.

\begin{figure}[t]
	\includegraphics[width=\columnwidth]{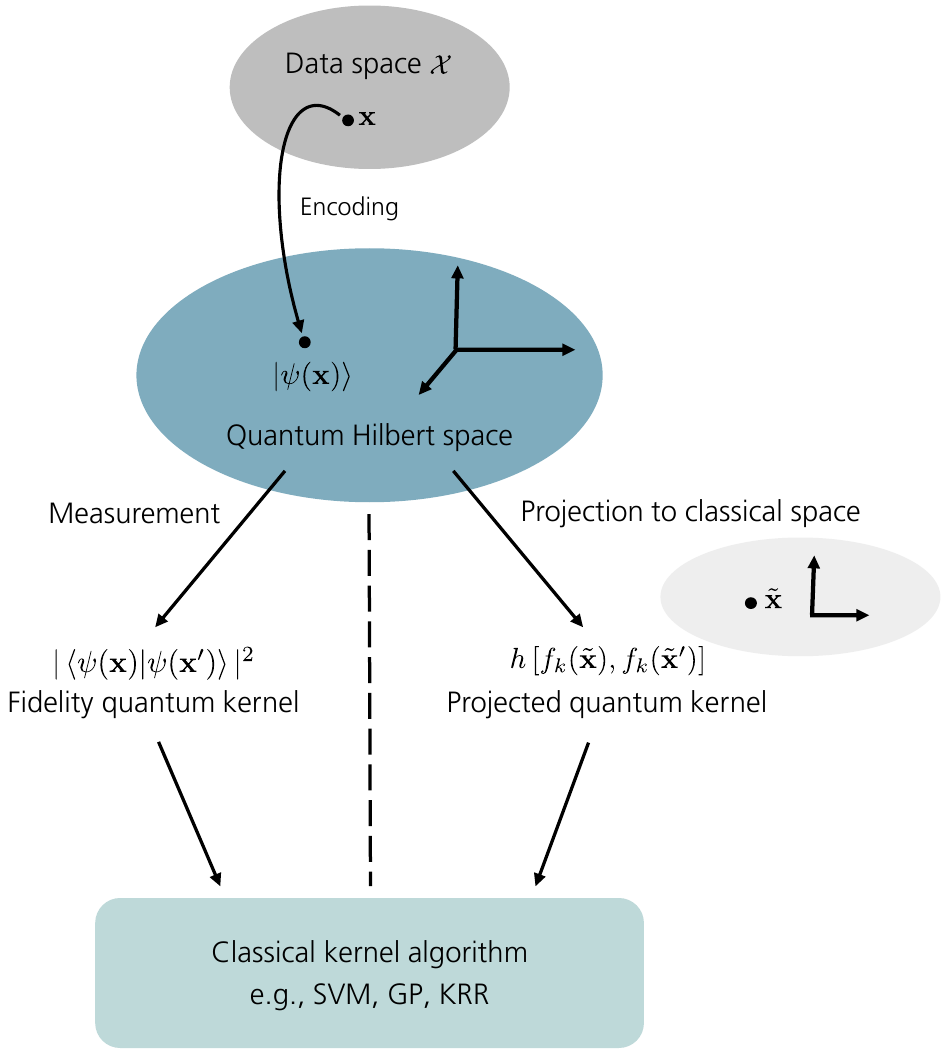}
	\caption{Schematic illustration of the function principle of quantum kernel methods, which can be formally embedded into the rich mathematical framework of conventional kernel theory. Data points $\xdata$ are mapped to the quantum Hilbert space by encoding them into a quantum state $\ket{\psi(\xdata)}$. Access to the quantum Hilbert space is granted by measurements. \textbf{Left:} In quantum mechanics, the straightforward way to express measurements is given by the Hilbert-Schmidt inner product. Quantum kernels which are defined using this fidelity-type metric are thus referred to as fidelity quantum kernels (FQKs). For pure states this reduces to Eq.~\eqref{eq:fidelity_quantum_kernel}. \textbf{Right:} When projecting the quantum states to an approximate classical representation using, e.g., reduced physical observables this gives rise to a family of projected quantum kernels (PQKs). One of the simplest forms of defining PQKs is to measure $k$-particle reduced density matrices, cf. Eq.~\eqref{eq:kRDM}, and process the result with a classical kernel function $h$. The resulting quantum kernel matrices (either FQK or PQK) are subsequently passed to a classical kernel algorithm such as a SVM or a GP.}
	\label{fig:kernel_sketch}
\end{figure}

In quantum computing, access to the quantum Hilbert space is given by measurements, which can also be expressed by inner products of quantum states, cf. Fig.~\ref{fig:kernel_sketch}. Thus, the kernel function can be defined using the native geometry of the quantum state space: the Hilbert-Schmidt inner product $\tr[\rho(\xdata)\rho(\xdata')]$. For pure states this reduces to
\begin{equation}
    \label{eq:fidelity_quantum_kernel}
    K^{\rm FQK}(\xdata, \xdata')=\abs{\braket{\psi(\xdata)|\psi(\xdata')}}^2\,,
\end{equation}
which, as it is based on a fidelity-type metric, is often called the fidelity quantum kernel (FQK)~\cite{Huang2021}. Rewriting $K^{\rm FQK}(\xdata, \xdata')=\abs{\bra{0}U^\dag(\xdata)U(\xdata')\ket{0}}^2$ reveals that in practise every Gram matrix element can be obtained by measuring the zero state after applying the quantum feature map and its Hermitian conjugate ~\cite{Jerbi2023}. 

With increasing problem size, FQKs can potentially suffer from exponential concentration leading to quantum models that can become untrainable~\cite{thanasilp2022}. To alleviate this problem, Ref.~\cite{Huang2021} introduced a family of projected quantum kernels (PQK) which project the quantum states to an approximate classical representation by using, e.g., reduced physical observables, cf. Fig.~\ref{fig:kernel_sketch}. A PQK can be thought of defining features in a classical vector space by taking a detour through a quantum Hilbert space. The result can be hard to compute without a quantum computer due to the quantum detour but still retains desirable properties of the classical feature space. They thus have several appealing properties compared to FQKs.

A simple way of defining a PQK is based on measuring $k$-particle reduced density matrices ($k$-RDMs),
\begin{equation}
    \label{eq:kRDM}
    \rho_{S_k}(\xdata) = \mathrm{tr}_{j\notin S_k}\left[\rho(\xdata)\right]\,,
\end{equation}
where $\mathrm{tr}_{j\notin S_k}$ is the partial trace over all qubits not in a subset $S_k$ of $k$ qubits. The result is then processed in a classical kernel function $h$. Since Eq.~\eqref{eq:kRDM} just depends on one data point at a time and the pair-wise Gram matrix is only calculated classically, for a proper choice of $h$, PQKs admit a linear scaling in terms of the needed quantum computing resources.

Measuring $k$-RDMs with respect to some observable, corresponds to evaluating a QNN for a $k$-local cost operator $O_k$. Thus, in sQUlearn, PQKs are generally defined as
\begin{equation}
    \label{eq:PQK-general}
    K^{\rm PQK}(\xdata, \xdata^\prime) = h\left[f_k(\xdata), f_k(\xdata^\prime)\right]\,,
\end{equation}
where $f_k$ is given in Eq.~\eqref{eq:quantum-model-general} with a local observable $O_k$.

As an example, the default PQK implementation in sQUlearn is given by measuring the 1-RDM on all qubits with respect to all Pauli operators $P\in\lbrace \sigma_x, \sigma_y, \sigma_z\rbrace$, i.e.,
\begin{equation}
    K^{\rm PQK}(\xdata,\xdata^\prime)=\exp\left(-\gamma\sum_{k,P}\left\lbrace\tr[P\rho_k(\xdata\boldsymbol)] - \tr[P\rho_k(\xdata^\prime)]\right\rbrace^2\right)\,,\label{eq:gaussian_pqk}
\end{equation}
where $\gamma\in\mathbb{R_+}$ is a hyperparameter and $\rho_k(\xdata)=\mathrm{tr}_{j\neq k}\left[\rho(\xdata)\right]$ is the 1-RDM for qubit $k$, i.e., the partial trace of the quantum state $\rho(\xdata)$ over all qubits except for the $k$-th qubit. This results in an RBF-like structure of the kernel.
 
\subsubsection{\label{subsubsec:kernel-implementation}Implementation}
\begin{figure}
	\includegraphics[width=\columnwidth]{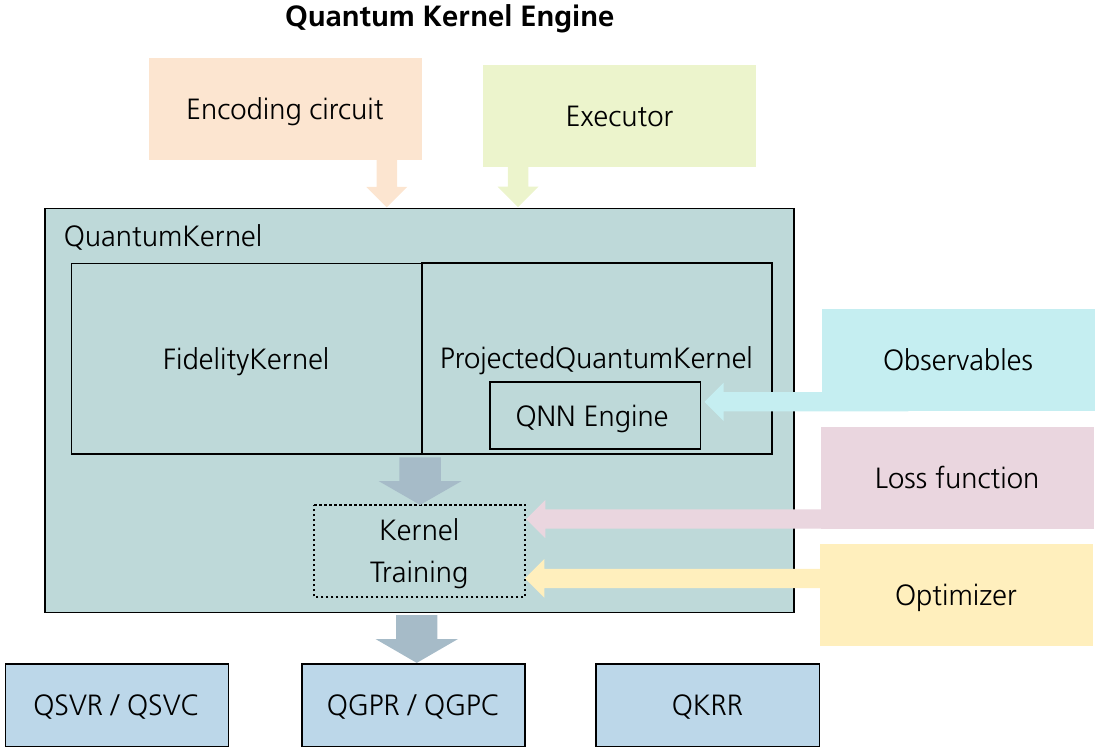}
	\caption{Schematic overview of sQUlearn's quantum kernel engine implementation including all dependencies on other program parts. To define a quantum kernel, the user first needs to provide an encoding circuit object which is used for embedding data into the quantum Hilbert space (cf. Info Box~\ref{box:encoding_circuits}). After defining a corresponding circuit, one can either define a \texttt{FidelityKernel}, or a \texttt{ProjectedQuantumKernel}. The latter makes use of the QNN Engine, cf. Fig.~\ref{fig:qnn_design} and the \texttt{Observables} class. With the Executor the Gram matrix can be evaluated on real quantum computers or a simulator backend with  automatic session handling. If the encoding circuit also contains trainable parameters, sQUlearn provides optimization routines to optionally optimize these parameters. With the kernel object created, or stored in from a previous calculation the high level methods (blue) can be initialized and executed (cf. Info Box~\ref{box:kernels}).}
	\label{fig:qkm_design}
\end{figure}

The quantum kernel engine of sQUlearn is designed and implemented as schematically shown in Fig.~\ref{fig:qkm_design}. In terms of high-level kernelized ML methods, sQUlearn provides several conventional kernel algorithms based on quantum kernel matrices for solving regression and classification problems. These methods come with an analogue easy-to-use interface like the scikit-learn estimators as demonstrated in Info Box~\ref{box:kernels}.

The user can choose between FQKs and PQKs which can be constructed from various (parametrized) data encoding circuits and which can be computed on either simulators or real backends via the \texttt{Executor} class (cf. Sec~\ref{sec:executor}). 
FQKs are realized in two distinct manners. If a statevector is available during simulation, the fidelity between states $\psi(\xdata)$ and $\psi(\xdata')$ is computed by numerically calculating the overlap of the statevectors. Otherwise, the kernel entries are computed using the fidelity test circuit, which measures the zero state of the circuit $U^\dagger(x')U(x)\ket{0}$.
For Qiskit backends, we utilize the quantum kernel implementations provided by Qiskit Machine Learning, and in case of a PennyLane device, a custom implementation of the fidelity kernel is available.
PQKs are implemented by employing sQUlearn's low-level Quantum Neural Network (QNN) implementation to evaluate $k$-local cost operators, as depicted in Eq.~\eqref{eq:PQK-general}. Here, the QNN results serve as features for an outer kernel. The default PQK setting within sQUlearn is given in Eq.~\eqref{eq:gaussian_pqk} but in addition, there are is a wide selection of additional outer kernels available and cost operators can be customized. A corresponding example is given in Appendix~\ref{sec:appendix_pqk}, demonstrating how to define a custom cost operator as well as changing the outer kernel.

Both fidelity and projected kernels can be used with quantum feature maps that include trainable parameters yielding quantum kernels that become variationally trainable. They can be optimized using several optimization strategies. sQUlearn offers loss functions including kernel target alignment and negative-log-likelihood, along with several optimizers such as Adam and SLSQP. Beyond that, the composition of multiple quantum kernels is possible, allowing for tailored solutions.

The following example illustrates the variational quantum kernel training procedure; assuming some training data (\texttt{x\_train} and \texttt{y\_train}) have to be previously defined.
\begin{lstlisting}[style=python]
feature_map = ChebyshevPQC(
    num_qubits=4, num_features=1, num_layers=1)
q_kernel = ProjectedQuantumKernel(
    feature_map, Executor())

target_alignment = TargetAlignment(
    quantum_kernel=q_kernel)
optimizer = KernelOptimizer(
    loss=target_alignment,
    optimizer=Adam(options={"maxiter": 20, "lr": 0.1}))

q_krr = QKRR(q_kernel)
q_krr.fit(x_train, y_train)
q_krr.predict(x_train)
\end{lstlisting}
In this example, a PQK is trained using the target alignment~\cite{Kandola2003,Cristianini2006,Hubregtsen2022} as a loss function and Adam as an optimizer. The kernel with the optimized parameters is then used for a KRR. The code assumes the availability of suitable training and test data. The full executable code can be found in Appendix~\ref{sec:appendix_kernel_optimization}.

Kernel matrices that are calculated on noisy backends (real quantum hardware or noisy simulators) are prone to lose their positive semi-definiteness. To preserve this necessary property, both the \texttt{FidelityKernel} and the \texttt{ProjectedQuantumKernel} implementation feature regularization techniques such as thresholding or Tikhonov-regularization~\cite{Wang2021, Hubregtsen2022}. For FQKs, sQUlearn features an additional option for mitigating depolarizing noise according to Ref.~\cite{Hubregtsen2022}. Appendix~\ref{sec:appendix_kernel_regularization_mitigation} provides an example showing how these options can be used to improve results obtained in the presence of noise.

\begin{infobox}{Quantum Kernel Methods}{kernels}
    
    NISQ-based quantum kernel methods compute the Gram matrix with a quantum computer and pass it to a conventional kernel algorithm such as an SVM. sQUlearn offers several out-of-the-box quantum kernel methods such as quantum SVMs for classifciation (QSVC) and regression (QSVR), quantum Gaussian processes (QGP) for classification (QGPC) and regression (QGPR) as well as a quantum kernel ridge regression routine (QKRR). The user can choose between FQKs and PQKs, see Eqs.~\eqref{eq:fidelity_quantum_kernel} and~\eqref{eq:PQK-general}, respectively. Setting up a quantum kernel method can be done with just a few lines of code
    \begin{lstlisting}[style=python]
from squlearn.encoding_circuit import (
    YZ_CX_EncodingCircuit)
from squlearn.kernel import FidelityKernel, QSVC
from squlearn.util import Executor

feature_map = YZ_CX_EncodingCircuit(
    num_qubits=5, num_features=2, num_layers=2)
q_kernel = FidelityKernel(
    feature_map, executor=Executor())
model = QSVC(q_kernel)
     \end{lstlisting}
    This initializes a QSVC using a FQK as defined in Eq.~\eqref{eq:fidelity_quantum_kernel}.
    To use a PQK instead, the \texttt{FidelityKernel} class can be exchanged with a \texttt{ProjectedQuantumKernel} class. The resulting estimator can then be used like any scikit-learn estimator, e.g. fitting data with \texttt{model.fit()} and predicting with \texttt{model.predict()}.
\end{infobox}

\subsection{Backend Integration \label{sec:executor}}
sQUlearn provides a simple and unified approach for executing QML tasks on real quantum computers or simulators.
This is achieved by centralizing all quantum job executions into a single module: the \texttt{Executor} class.
It enables users to execute longer quantum programs without the need for constant monitoring and supervision, since it takes care of resubmitting failed jobs and automatically re-initializing expired sessions.
In the case of stochastic simulations with finite-sampling, a seed can be set at the beginning to provide reproducible results.
Results from remote quantum computations are automatically cached locally and can be reloaded when necessary.
This feature proves especially useful in cases of unexpected interruptions that require a restart, for example a training of a QNN. Event logging offers insights into background processes during longer runs. 

The Executor supports both, Qiskit and PennyLane backends for simulators or real hardware access. 
Transitioning between different backends within sQUlearn is as simple as modifying the Executor initialization, with all other operations, such as Qiskit Primtive creation or PennyLane circuit conversion, handled automatically in the background.
Consequently, this strongly simplifies the process of prototyping QML applications, since transitioning from a local simulator environment to real quantum hardware is accomplished with minimal adjustments, leaving the rest of the program untouched.
Additionally, users can choose from various other options such as utilizing a pre-configured Qiskit Runtime primitive, or selecting a backend from alternative providers that offer Qiskit or PennyLane implementations.

Internally, the Qiskit functionality is achieved by providing Primitives that override Qiskit's Sampler and Estimator primitives and route every execution through the Executor.
Since these primitives are derived from Qiskit, they can be utilized in other parts of Qiskit as well.
It's important to emphasize that this overwritten Primitives are not limited to QML applications alone; they can also be employed in other quantum programs, such as variational quantum eigensolver~\cite{Peruzzo2014} or similar programs derived from Qiskit's Sampler and Estimator primitives.

In the following example, we demonstrate the necessary steps to run the code provided in Info Box~\ref{box:kernels} on actual quantum hardware, specifically using the IBM quantum backend \texttt{ibm\_kyoto}.
\begin{lstlisting}[style=python]   
service = QiskitRuntimeService(channel="ibm_quantum")
executor = Executor(service.get_backend("ibm_kyoto"))
executor.set_primitive_options(resilience_level=1)

feature_map = YZ_CX_EncodingCircuit(
    num_qubits=3, num_features=1, num_layers=2)
q_kernel = FidelityKernel(feature_map, executor=executor)
model = QSVR(q_kernel)
\end{lstlisting}
First, a connection to the IBM Quantum service is established. It is assumed that the user has previously set up access to the IBM Quantum platform.\footnote{Refer to Ref.~\cite{qiskitruntime} for instructions.} Subsequently, the Executor is initialized with the backend obtained from the Qiskit service. The following steps, such as session creation and primitive initialization, are handled internally within the Executor.
Error mitigation can be configured directly in the settings of the Executor. The remaining part of the example is analogous to Info Box~\ref{box:kernels}.
The executable code including the imports is presented in Appendix~\ref{sec:appendix_executor_example}.
The example illustrates how sQUlearn enables a effortless transition from the simulation environment to the real backend, requiring only changes in the Executor initialization.

\section{Example: Creating a ML pipeline}
\label{sec:example}

In this example, we demonstrate how sQUlearn based algorithms can be embedded in standard ML workflows by performing a architecture search and hyperpameter optimization using scikit-learn methods. We illustrate this by optimizing a QSVM over several feature maps, circuit parameters and classical hyperparameters of the SVM. 

We start by creating a classification dataset with $100$ data points and $10$ features which are split into training and test sets with a ratio of $0.2$.
\begin{lstlisting}[style=python]
from sklearn.datasets import make_classification
from sklearn.model_selection import train_test_split

X, y = make_classification(
    n_samples=100, n_features=10, random_state=0)
X_train, X_test, y_train, y_test = train_test_split(
    X, y, test_size=0.2, random_state=0)
\end{lstlisting}
Next, we initialise a QSVC using a PQK with the high-dimensional encoding circuit shown in Info Box~\ref{box:encoding_circuits}.
\begin{lstlisting}[style=python]
from squlearn import Executor
from squlearn.encoding_circuit import (
    HighDimEncodingCircuit)
from squlearn.kernel import ProjectedQuantumKernel, QSVC

high_dim_ec = HighDimEncodingCircuit(
    num_qubits=4, num_features=3, num_layers=2)

pqk = ProjectedQuantumKernel(
    encoding_circuit=high_dim_ec,
    executor=Executor())

qsvc = QSVC(quantum_kernel=pqk)
\end{lstlisting}
Most of the common encoding schemes require a certain scaling in order to preserve injectivity of the embedding. To automatically handle this, we include the QSVC in a scikit-learn pipeline that preprocesses the data with a MinMaxScaler that normalizes the data between $0$ and $1$. Today's QML methods performed on NISQ computers are often limited regarding the dimensionalty of the dataset. We can account for by including a feature selection algorithm in the pipeline, reducing the number of features from $10$ to $3$. The resulting pipeline automatically performs all necessary preprocesing, enabling the QSVC to be applied to a wide range of datasets.
\begin{lstlisting}[style=python]
from sklearn.feature_selection import SelectKBest
from sklearn.pipeline import Pipeline
from sklearn.preprocessing import MinMaxScaler

scaler = MinMaxScaler(feature_range=(0.01, 0.99))
feature_selection = SelectKBest(k=3)
pipeline = Pipeline([
    ("scaler", scaler),
    ("feature_selection", feature_selection),
    ("qsvc", qsvc)])
\end{lstlisting}
Since the encoding circuit and its structure is one of the most important properties of a quantum kernel method, we benchmark the \texttt{HighDimEncodingCircuit} against a different encoding circuit. To demonstrate sQUlearn's capabilities, we do not resort to a predefined encoding circuit but rather define one from scratch using the \texttt{LayeredEncodingCircuit}. This feature allows us to create encoding circuits using string-based input which is shown in the following:
\begin{lstlisting}[style=python]
from squlearn.encoding_circuit import (
    LayeredEncodingCircuit)

layered_ec = LayeredEncodingCircuit.from_string(
    "Ry(x)-Rz(x)-Rx(p)-cx",
    num_qubits=4,
    num_features=3,
    num_layers=2)
\end{lstlisting}
Here, \texttt{Rx}, \texttt{Ry} and \texttt{Ry} are rotation gates where each gate in the string is applied to all the qubits, and therefore, forms a layer (hence the name \texttt{LayeredEnodingCircuit}).  
The symbols \texttt{x} and \texttt{p} indicate features and trainable parameters, respectively. 
The number of features is set to three since the dimensionality is reduced to three in the scikit-learn pipeline by the feature selection.
The automatic mapping of feature vector components to qubits is managed by the implementation, with each feature being placed sequentially in various gates, one at a time.
This process occurs in a loop, ensuring that the placement is repeated after the last component is reached.
The trainable parameters are considered to be different for every single gate, and the dimension of parameter vector is determined automatically.
The layered encoding circuit supports not only the displayed rotation and CNOT gates but also various other one- and two-qubit gates. Furthermore, it allows for custom non-linear input encoding functions and different layers.
For more detailed information on the construction, please refer to the sQUlearn documentation \cite{squlearnDocu2023}.

Having set up the pipeline and the encoding circuits which we want to compare, we now initialize a grid search to identify the optimal circuit configuration within a predefined search space. In this search, we not only include two different encoding circuits but also optimize over the number of qubits and layers, the outer kernel of the PQK as well as the classical hyperparameters of the SVM in a single operation.
\begin{lstlisting}[style=python]
from sklearn.model_selection import GridSearchCV

param_grid = {
    "qsvc__encoding_circuit": [layered_ec, high_dim_ec],
    "qsvc__outer_kernel": ["Gaussian", "DotProduct"],
    "qsvc__num_qubits": [3, 4],
    "qsvc__num_layers": [1, 2, 3],
    "qsvc__C": [0.01, 0.1, 1.0]}

grid_search = GridSearchCV(pipeline, param_grid, cv=5)
grid_search.fit(X_train, y_train)
\end{lstlisting}
In the code above, the prefix \texttt{qsvc\_\_} is required when using a scikit-learn pipeline instead of an estimator directly. 
The example returns the trained QSVC with an optimal configuration of the encoding circuit and the SVM within the defined search space. 
Similar hyperparameter optimizations, as demonstrated in the example, can be extended to all sQUlearn high-level methods. The effortless integration into established ML workflows not only simplifies but also enhances the development and optimization of intricate QML models, and it is one of the unique features of sQUlearn.

\section{Conclusion}
\label{sec:conclusion}

In this work, we have introduced sQUlearn, an open-source Python library aimed at simplifying and streamlining the development of QML models. Designed for compatibility with existing ML tools, sQUlearn offers a scikit-learn-like interface, enabling seamless integration into a broad array of pipelines and frameworks.
Our library targets both practitioners and researchers, offering high-level implementations for various state-of-the-art algorithms and low-level features for advanced development.
It provides a comprehensive suite of tools designed for effortless creation and training of custom QML models that leverage the power of QNNs and quantum kernels.
Leveraging the capabilities of both Qiskit and PennyLane, sQUlearn provides a diverse array of execution options across various simulated and real quantum hardware backends.

The version described in this paper is sQUlearn 0.7.0. For comprehensive documentation, sample use-cases, and installation guidelines, we direct readers to the sQUlearn website~\cite{squlearnDocu2023}. As with any initial software release, in particular in a rapidly evolving field such as QML, expect the capabilities and performance of sQUlearn to expand over time, responding to the changing needs and discoveries in both the quantum and classical ML communities. We thus anticipate future versions of sQUlearn to incorporate advancements such as optimization techniques tailored for NISQ devices, access to various more quantum hardware and simulators, data-preprocessing designed for QML algorithms, and finally, additional QML models that leverage quantum computing capabilities. 

\section*{Code availability}
sQUlearn is an open-source software project licensed under the Apache License 2.0. The source code is accessible on GitHub.\footnote{\url{https://github.com/squlearn/squlearn}} Additional tutorials, instructions, and an installation guide can be accessed through the documentation.\footnote{\url{https://squlearn.github.io}}
All shown examples have been written and tested with sQUlearn 0.7.0.

\begin{acknowledgments}
This work was supported by the German Federal Ministry of Economic Affairs and Climate Action through the projects AutoQML (grant no.~01MQ22002A) and AQUAS (grant no.~01MQ22003D), as well as the German Federal Ministry of Education and Research through the project H2Giga $\text{Degrad-EL}^3$ (grant no.~03HY110D). The authors would like to thank André Kempinger for his contributions to sQUlearn.
\end{acknowledgments}

\appendix

\section{Quantum Neural Network Examples}

\subsection{QNN Regression Example \label{sec:qnn_example}}

\begin{figure}
\centering
\includegraphics[width=8cm]{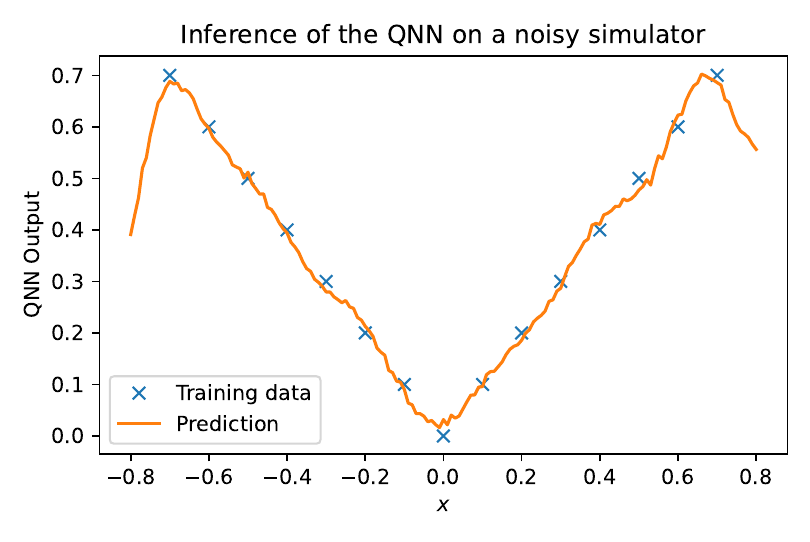}\hspace*{1.2cm}
\caption{Output of the QNN example (cf. Sec. \ref{sec:qnn_example}) \label{fig:qnn_inference} }
\end{figure}

The following examples displays the training and inference of a QNN (cf. example in Sec. \ref{sec:qnn_implementation}) by curve fitting the absolute value function with a shot-based simulator. 
Furthermore, the code demonstrates the set-up of the variance regularization and an automatic shot adjustment. 
The output of the printed figure is displayed in Fig.~\ref{fig:qnn_inference}.

\label{section:appendix_qnn_example}
\begin{lstlisting}[style=python]
import numpy as np
import matplotlib.pyplot as plt

from squlearn import Executor
from squlearn.encoding_circuit import ChebyshevPQC
from squlearn.observables import SummedPaulis
from squlearn.qnn import (
    QNNRegressor,
    SquaredLoss,
    ShotsFromRSTD)
from squlearn.optimizers import Adam

from qiskit.primitives import Sampler

# Initilize encoding circuit
encoding_circuit = ChebyshevPQC(
    num_qubits=4, num_features=1, num_layers=3)

# Initilize observable
observable = SummedPaulis(num_qubits=4)

# Set-up QNN regression
qnn = QNNRegressor(
    encoding_circuit,
    observable,
    Executor(Sampler(),shots=5000),
    SquaredLoss(),
    Adam(options={"lr": 0.3}),
    variance=0.005,
    shot_control=ShotsFromRSTD()
    )

# Set training data
x_train = np.arange(-0.7, 0.8, 0.1)
y_train = np.abs(x_train)

# Train the model
qnn.fit(x_train, y_train)

# Data for testing the trained model
x_test = np.arange(-0.8, 0.81, 0.01)
y_test = qnn.predict(x_test)

# Generate a plot of the QNN inference
plt.plot(x_train, y_train, "x", label="Training data")
plt.plot(x_test, y_test, label="Prediction")
plt.title("Inference of the QNN on a noisy simulator")
plt.xlabel("$x$")
plt.ylabel("QNN Output")
plt.legend()
plt.show()
\end{lstlisting}

\subsection{QCNN Classification Example \label{sec:appendix_qcnn_example}}
The following code shows a minimal classification example for setting up and training a QCNN including mid-circuit measurements with sQUlearn. 

\begin{lstlisting}[style=python]
import numpy as np

from sklearn.datasets import make_moons
from sklearn.model_selection import train_test_split

from squlearn.encoding_circuit import QCNNEncodingCircuit
from squlearn.qnn import QNNClassifier, SquaredLoss
from squlearn.util import Executor
from squlearn.optimizers import Adam

# Generate data from the make_moons dataset
X, y = make_moons(
    noise=0.3, random_state=1, n_samples=100)
X_train, X_test, y_train, y_test = train_test_split(
    X, y, test_size=0.3, random_state=0)

# Initialize default QCNN with 4 qubits with one 
# convolutional layer, one pooling layer with
# mid-circuit measurements and one fully connected layer
qc = QCNNEncodingCircuit(4, 2)
qc.convolution()
qc.pooling(measurement=True)
qc.repeat_layers()
qc.fully_connected()

# Initialize default QCNN observable
obs = qc.QCNNObservable()

# Set-up QNN classification
qnn = QNNClassifier(
    qc, obs, Executor(), SquaredLoss(), Adam())

# Train the model
qnn.fit(X_train, y_train)

# Data for testing the trained model
print("Training score:", qnn.score(X_train, y_train))
print("Testing score:", qnn.score(X_test, y_test))
\end{lstlisting}

\section{Quantum Kernel Examples}
\label{sec:appendix_kernel_example}
This section provides the complete code for the examples referenced in the main text in Sec.~\ref{subsubsec:kernel-implementation}.

\subsection{Projected Quantum Kernel Example}
\label{sec:appendix_pqk}

\begin{figure}
\centering
\includegraphics[width=8cm]{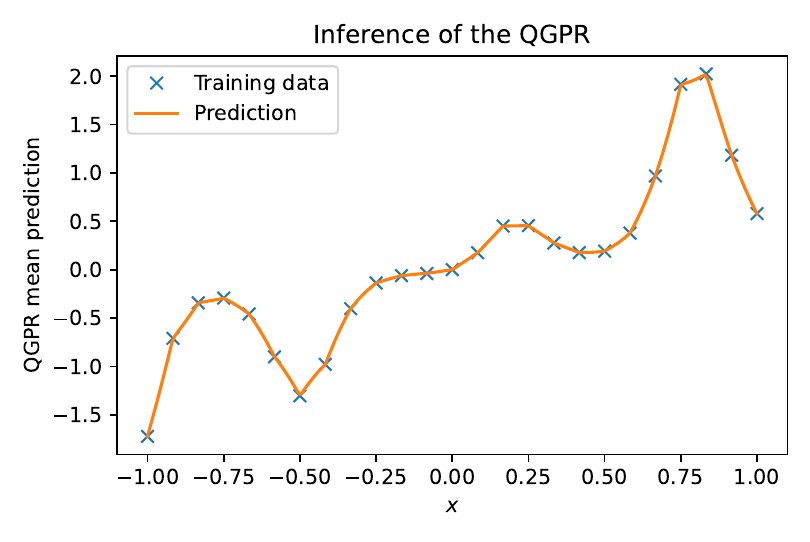}\hspace*{1.2cm}
\caption{\label{fig:pqk_result}Output of the PQK example.}
\end{figure}

The following example shows how to change the measurement operators for evaluating expectation values in PQK calculations. Moreover, the following code demonstrates how to change the functional form of the outer kernel. Gaussian Process regression is used as estimator. The results are shown in Fig.~\ref{fig:pqk_result}.
\begin{lstlisting}[style=python]
import numpy as np
import matplotlib.pyplot as plt
from squlearn.encoding_circuit import (
    YZ_CX_EncodingCircuit)
from squlearn.observables import CustomObservable
from squlearn.kernel import (
    ProjectedQuantumKernel,
    QGPR)
from squlearn.util import Executor


# Generate training & test data
def func(x):
    return x * np.exp(np.sin(10 * x))


x_train = np.linspace(-1, 1, 25)
y_train = func(x_train)
x_train = x_train.reshape(-1, 1)
x_test = np.linspace(-1, 1, 200)
y_test = func(x_test)
x_test = x_train.reshape(-1, 1)

feature_map = YZ_CX_EncodingCircuit(
    num_qubits=4, num_features=1, num_layers=2)

# Create custom observables
measurements = []
measurements.append(CustomObservable(4, "ZZZZ"))
measurements.append(CustomObservable(4, "YYYY"))
measurements.append(CustomObservable(4, "XXXX"))

# Use Matern outer kernel with nu=0.5
# as a outer kernel hyperparameter
kernel = ProjectedQuantumKernel(
    feature_map,
    executor=Executor(),
    measurement=measurements,
    outer_kernel="matern",
    nu=0.5)

q_gpr = QGPR(kernel)
q_gpr.fit(x_train, y_train)
y_pred = q_gpr.predict(x_test)

plt.plot(x_train, y_train, "x", label="Training data")
plt.plot(x_test, y_pred, label="Prediction")
plt.title("Inference of the QGPR")
plt.xlabel("$x$")
plt.ylabel("QGPR mean prediction")
plt.legend()
plt.show()
\end{lstlisting}

\subsection{Quantum Kernel Optimization Example}
\label{sec:appendix_kernel_optimization}

\begin{figure}
\includegraphics[width=8cm]{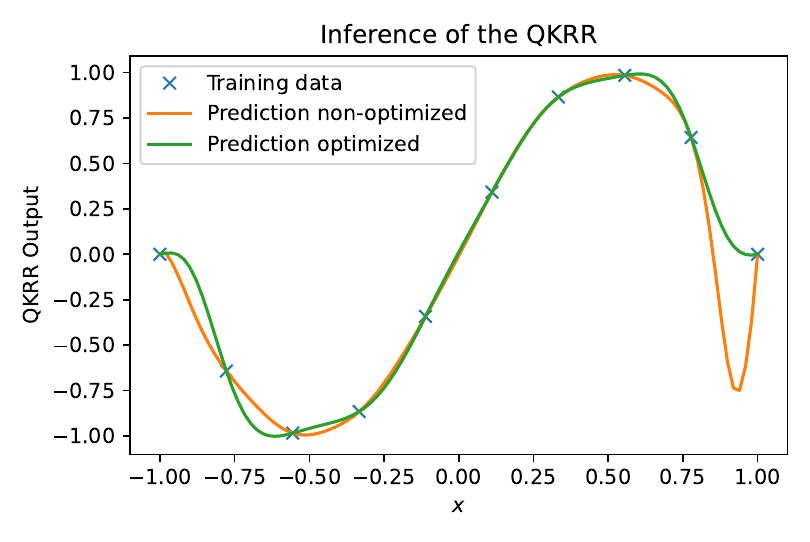}\hspace*{1.2cm}
\caption{Output of the quantum kernel optimization example (cf. Sec. \ref{subsubsec:kernel-implementation} \label{fig:qkm_result} }
\end{figure}

The following provides the complete code for running the kernel optimization example introduced in Sec.~\ref{subsubsec:kernel-implementation}. The results can be seen in Fig.~\ref{fig:qkm_result}

\begin{lstlisting}[style=python]
import matplotlib.pylab as plt
import numpy as np

from squlearn.encoding_circuit import ChebyshevPQC
from squlearn.kernel import (
    ProjectedQuantumKernel,
    QKRR)
from squlearn.kernel.optimization import (
    KernelOptimizer,
    TargetAlignment)
from squlearn.optimizers import Adam
from squlearn.util import Executor

x_train = np.linspace(-1, 1, 10).reshape(-1, 1)
x_test = np.linspace(-1, 1, 100).reshape(-1, 1)
y_train = np.sin(x_train.ravel() * np.pi)

feature_map = ChebyshevPQC(
    num_qubits=4, num_features=1, num_layers=1)
q_kernel = ProjectedQuantumKernel(
    feature_map, Executor())

q_krr = QKRR(q_kernel)
q_krr.fit(x_train, y_train)
prediction_untrained = q_krr.predict(x_test)

target_alignment = TargetAlignment(
    quantum_kernel=q_kernel)
optimizer = KernelOptimizer(
    loss=target_alignment,
    optimizer=Adam(options={"maxiter": 20, "lr": 0.1}))
optimizer.run_optimization(X=x_train, y=y_train)

q_krr = QKRR(q_kernel)
q_krr.fit(x_train, y_train)
prediction = q_krr.predict(x_test)

plt.plot(x_train, y_train, "x", label="Training data")
plt.plot(
    x_test,
    prediction_untrained,
    label="Prediction non-optimized")
plt.plot(
    x_test, prediction, label="Prediction optimized")
plt.title("Inference of the QKRR")
plt.xlabel("$x$")
plt.ylabel("QKRR Output")
plt.legend()
plt.show()
\end{lstlisting}

\subsection{Quantum Kernel Regularization and Mitigation Example}
\label{sec:appendix_kernel_regularization_mitigation}
In general, the concept of regularization is crucial for quantum kernel methods. This importance arises from the fact that kernels calculated on quantum computers (or simulators) can lose their positive semi-definiteness when exposed to noise.
However, positive semi-definiteness is a central and necessary property in kernel theory (cf. Sec.\ref{subsubsec:kernel-background}). To uphold this essential characteristic, both the \texttt{FidelityKernel} and the \texttt{ProjectedQuantumKernel} implementations in sQUlearn incorporate the two regularization techniques outlined in Ref.~\cite{Hubregtsen2022}.
Moreover, when computing Fidelity Quantum Kernels (FQKs) on real backends, sQUlearn includes an option to mitigate depolarizing noise, as outlined in Ref.~\cite{Hubregtsen2022}. The formula for this mitigation technique is derived by leveraging the observation that, in the ideal scenario, all diagonal elements of training kernel matrices equal one. While this property is inherently fulfilled for Projected Quantum Kernels (PQKs) by construction, it is often violated for FQKs.

The utilization of both regularization and the depolarizing noise mitigation option is exemplified in the code below. Regularization is illustrated by printing the corresponding eigenvalues, while mitigation is showcased through the display of diagonal elements in the kernel matrices.
\begin{lstlisting}[style=python]
import numpy as np
from qiskit_ibm_runtime.fake_provider import FakeManilaV2
from squlearn.encoding_circuit import (
    YZ_CX_EncodingCircuit)
from squlearn.kernel import FidelityKernel
from squlearn.util import Executor

# Generate some data to evaluate a matrix
x = np.linspace(-1.5, 1.5, 20)

enc_circ = YZ_CX_EncodingCircuit(
    num_qubits=5, num_features=1, num_layers=3)

# Define the executor with a fake backend 
# and set the desired number of shots
executor = Executor(FakeManilaV2(), shots=500)

# Compute the Kernel and print out the eigenvalues
FQK = FidelityKernel(
    enc_circ,
    executor=executor,
    evaluate_duplicates="all")
kernel_matrix_1 = FQK.evaluate(x.reshape(-1, 1))
print("Eigenvalues without regularization: ",
       np.linalg.eigvalsh(kernel_matrix_1))

# Switch to Tikhonov regularization and print out the
# eigenvalues for comparison
FQK.set_params(regularization="tikhonov")
kernel_matrix_2 = FQK.evaluate(x.reshape(-1, 1))
print("Eigenvalues with regularization: ",
       np.linalg.eigvalsh(kernel_matrix_2))

# We now show how to mitigate depolarization noise
# in the kernel matrix
FQK_mitigated = FidelityKernel(
    enc_circ,
    executor=executor,
    evaluate_duplicates="all",
    mit_depol_noise="msplit")
kernel_matrix_3 = FQK_mitigated.evaluate(
    x.reshape(-1, 1))
print("The diagonal of the unmitigated matrix is: ",
       np.diag(kernel_matrix_1))
print("The diagonal of mitigated matrix is: ",
       np.diag(kernel_matrix_3))
\end{lstlisting}

\section{Executor Example \label{sec:appendix_executor_example}}

The following example demonstrates the configuration of the Executor to execute the code provided in Info Box \ref{box:kernels} on actual quantum hardware (refer to the example in Sec. \ref{sec:executor}). Furthermore, it includes the setup of options for error mitigation.

\begin{lstlisting}[style=python]
from qiskit_ibm_runtime import QiskitRuntimeService
from squlearn import Executor
from squlearn.encoding_circuit import (
    YZ_CX_EncodingCircuit)
from squlearn.kernel import FidelityKernel, QSVR

service = QiskitRuntimeService(channel="ibm_quantum")
executor = Executor(
    service.get_backend("ibm_kyoto"))
executor.set_primitive_options(resilience_level=1)

feature_map = YZ_CX_EncodingCircuit(
    num_qubits=3, num_features=1, num_layers=2)
q_kernel = FidelityKernel(
    feature_map, executor=executor)
model = QSVR(q_kernel)
\end{lstlisting}


\begin{thebibliography}{62}%
\makeatletter
\providecommand \@ifxundefined [1]{%
 \@ifx{#1\undefined}
}%
\providecommand \@ifnum [1]{%
 \ifnum #1\expandafter \@firstoftwo
 \else \expandafter \@secondoftwo
 \fi
}%
\providecommand \@ifx [1]{%
 \ifx #1\expandafter \@firstoftwo
 \else \expandafter \@secondoftwo
 \fi
}%
\providecommand \natexlab [1]{#1}%
\providecommand \enquote  [1]{``#1''}%
\providecommand \bibnamefont  [1]{#1}%
\providecommand \bibfnamefont [1]{#1}%
\providecommand \citenamefont [1]{#1}%
\providecommand \href@noop [0]{\@secondoftwo}%
\providecommand \href [0]{\begingroup \@sanitize@url \@href}%
\providecommand \@href[1]{\@@startlink{#1}\@@href}%
\providecommand \@@href[1]{\endgroup#1\@@endlink}%
\providecommand \@sanitize@url [0]{\catcode `\\12\catcode `\$12\catcode
  `\&12\catcode `\#12\catcode `\^12\catcode `\_12\catcode `\%12\relax}%
\providecommand \@@startlink[1]{}%
\providecommand \@@endlink[0]{}%
\providecommand \url  [0]{\begingroup\@sanitize@url \@url }%
\providecommand \@url [1]{\endgroup\@href {#1}{\urlprefix }}%
\providecommand \urlprefix  [0]{URL }%
\providecommand \Eprint [0]{\href }%
\providecommand \doibase [0]{https://doi.org/}%
\providecommand \selectlanguage [0]{\@gobble}%
\providecommand \bibinfo  [0]{\@secondoftwo}%
\providecommand \bibfield  [0]{\@secondoftwo}%
\providecommand \translation [1]{[#1]}%
\providecommand \BibitemOpen [0]{}%
\providecommand \bibitemStop [0]{}%
\providecommand \bibitemNoStop [0]{.\EOS\space}%
\providecommand \EOS [0]{\spacefactor3000\relax}%
\providecommand \BibitemShut  [1]{\csname bibitem#1\endcsname}%
\let\auto@bib@innerbib\@empty
%</preamble>
\bibitem [{\citenamefont {Chui}\ \emph {et~al.}(2023)\citenamefont {Chui},
  \citenamefont {Hazan}, \citenamefont {Roberts}, \citenamefont {Singla},\ and\
  \citenamefont {Smaje}}]{chui2023}%
  \BibitemOpen
  \bibfield  {author} {\bibinfo {author} {\bibfnamefont {M.}~\bibnamefont
  {Chui}}, \bibinfo {author} {\bibfnamefont {E.}~\bibnamefont {Hazan}},
  \bibinfo {author} {\bibfnamefont {R.}~\bibnamefont {Roberts}}, \bibinfo
  {author} {\bibfnamefont {A.}~\bibnamefont {Singla}},\ and\ \bibinfo {author}
  {\bibfnamefont {K.}~\bibnamefont {Smaje}},\ }\href@noop {} {\bibinfo {title}
  {The economic potential of generative ai}} (\bibinfo {year}
  {2023})\BibitemShut {NoStop}%
\bibitem [{\citenamefont {Samuel}(1959)}]{Samuel1956}%
  \BibitemOpen
  \bibfield  {author} {\bibinfo {author} {\bibfnamefont {A.~L.}\ \bibnamefont
  {Samuel}},\ }\bibfield  {title} {\bibinfo {title} {Some studies in machine
  learning using the game of checkers},\ }\href
  {https://doi.org/10.1147/rd.33.0210} {\bibfield  {journal} {\bibinfo
  {journal} {IBM Journal of Research and Development}\ }\textbf {\bibinfo
  {volume} {3}},\ \bibinfo {pages} {210} (\bibinfo {year} {1959})}\BibitemShut
  {NoStop}%
\bibitem [{\citenamefont {Crevier}(1993)}]{Crevier1993}%
  \BibitemOpen
  \bibfield  {author} {\bibinfo {author} {\bibfnamefont {D.}~\bibnamefont
  {Crevier}},\ }\href@noop {} {\emph {\bibinfo {title} {AI: the tumultuous
  history of the search for artificial intelligence}}}\ (\bibinfo  {publisher}
  {Basic Books, Inc.},\ \bibinfo {year} {1993})\BibitemShut {NoStop}%
\bibitem [{\citenamefont {Linnainmaa}(1976)}]{Linnainmaa1976}%
  \BibitemOpen
  \bibfield  {author} {\bibinfo {author} {\bibfnamefont {S.}~\bibnamefont
  {Linnainmaa}},\ }\bibfield  {title} {\bibinfo {title} {Taylor expansion of
  the accumulated rounding error},\ }\href {https://doi.org/10.1007/BF01931367}
  {\bibfield  {journal} {\bibinfo  {journal} {BIT Numerical Mathematics}\
  }\textbf {\bibinfo {volume} {16}},\ \bibinfo {pages} {146} (\bibinfo {year}
  {1976})}\BibitemShut {NoStop}%
\bibitem [{\citenamefont {Werbos}(1982)}]{Werbos1982}%
  \BibitemOpen
  \bibfield  {author} {\bibinfo {author} {\bibfnamefont {P.}~\bibnamefont
  {Werbos}},\ }\bibfield  {title} {\bibinfo {title} {Applications of advances
  in nonlinear sensitivity analysis},\ }\href@noop {} {\bibfield  {journal}
  {\bibinfo  {journal} {System modeling and optimization}\ ,\ \bibinfo {pages}
  {762–770}} (\bibinfo {year} {1982})}\BibitemShut {NoStop}%
\bibitem [{\citenamefont {Cerezo}\ \emph {et~al.}(2022)\citenamefont {Cerezo},
  \citenamefont {Verdon}, \citenamefont {Huang}, \citenamefont {Cincio},\ and\
  \citenamefont {Coles}}]{Cerezo2022}%
  \BibitemOpen
  \bibfield  {author} {\bibinfo {author} {\bibfnamefont {M.}~\bibnamefont
  {Cerezo}}, \bibinfo {author} {\bibfnamefont {G.}~\bibnamefont {Verdon}},
  \bibinfo {author} {\bibfnamefont {H.-Y.}\ \bibnamefont {Huang}}, \bibinfo
  {author} {\bibfnamefont {L.}~\bibnamefont {Cincio}},\ and\ \bibinfo {author}
  {\bibfnamefont {P.~J.}\ \bibnamefont {Coles}},\ }\bibfield  {title} {\bibinfo
  {title} {Challenges and opportunities in quantum machine learning},\ }\href
  {https://doi.org/10.1038/s43588-022-00311-3} {\bibfield  {journal} {\bibinfo
  {journal} {Nature Computational Science}\ }\textbf {\bibinfo {volume} {2}},\
  \bibinfo {pages} {567} (\bibinfo {year} {2022})}\BibitemShut {NoStop}%
\bibitem [{\citenamefont {Harrow}\ \emph {et~al.}(2009)\citenamefont {Harrow},
  \citenamefont {Hassidim},\ and\ \citenamefont {Lloyd}}]{Harrow2009}%
  \BibitemOpen
  \bibfield  {author} {\bibinfo {author} {\bibfnamefont {A.~W.}\ \bibnamefont
  {Harrow}}, \bibinfo {author} {\bibfnamefont {A.}~\bibnamefont {Hassidim}},\
  and\ \bibinfo {author} {\bibfnamefont {S.}~\bibnamefont {Lloyd}},\ }\bibfield
   {title} {\bibinfo {title} {Quantum algorithm for linear systems of
  equations},\ }\href {https://doi.org/10.1103/PhysRevLett.103.150502}
  {\bibfield  {journal} {\bibinfo  {journal} {Phys. Rev. Lett.}\ }\textbf
  {\bibinfo {volume} {103}},\ \bibinfo {pages} {150502} (\bibinfo {year}
  {2009})}\BibitemShut {NoStop}%
\bibitem [{\citenamefont {Rebentrost}\ \emph {et~al.}(2014)\citenamefont
  {Rebentrost}, \citenamefont {Mohseni},\ and\ \citenamefont
  {Lloyd}}]{Rebentrost2014}%
  \BibitemOpen
  \bibfield  {author} {\bibinfo {author} {\bibfnamefont {P.}~\bibnamefont
  {Rebentrost}}, \bibinfo {author} {\bibfnamefont {M.}~\bibnamefont
  {Mohseni}},\ and\ \bibinfo {author} {\bibfnamefont {S.}~\bibnamefont
  {Lloyd}},\ }\bibfield  {title} {\bibinfo {title} {Quantum support vector
  machine for big data classification},\ }\href
  {https://doi.org/10.1103/PhysRevLett.113.130503} {\bibfield  {journal}
  {\bibinfo  {journal} {Phys. Rev. Lett.}\ }\textbf {\bibinfo {volume} {113}},\
  \bibinfo {pages} {130503} (\bibinfo {year} {2014})}\BibitemShut {NoStop}%
\bibitem [{\citenamefont {Zhao}\ \emph {et~al.}(2019)\citenamefont {Zhao},
  \citenamefont {Fitzsimons},\ and\ \citenamefont {Fitzsimons}}]{Zhao2019}%
  \BibitemOpen
  \bibfield  {author} {\bibinfo {author} {\bibfnamefont {Z.}~\bibnamefont
  {Zhao}}, \bibinfo {author} {\bibfnamefont {J.~K.}\ \bibnamefont
  {Fitzsimons}},\ and\ \bibinfo {author} {\bibfnamefont {J.~F.}\ \bibnamefont
  {Fitzsimons}},\ }\bibfield  {title} {\bibinfo {title} {Quantum-assisted
  gaussian process regression},\ }\href
  {https://doi.org/10.1103/PhysRevA.99.052331} {\bibfield  {journal} {\bibinfo
  {journal} {Phys. Rev. A}\ }\textbf {\bibinfo {volume} {99}},\ \bibinfo
  {pages} {052331} (\bibinfo {year} {2019})}\BibitemShut {NoStop}%
\bibitem [{\citenamefont {Liu}\ \emph {et~al.}(2023)\citenamefont {Liu},
  \citenamefont {Liu}, \citenamefont {Liu}, \citenamefont {Ye}, \citenamefont
  {Wang}, \citenamefont {Alexeev}, \citenamefont {Eisert},\ and\ \citenamefont
  {Jiang}}]{Liu2023}%
  \BibitemOpen
  \bibfield  {author} {\bibinfo {author} {\bibfnamefont {J.}~\bibnamefont
  {Liu}}, \bibinfo {author} {\bibfnamefont {M.}~\bibnamefont {Liu}}, \bibinfo
  {author} {\bibfnamefont {J.-P.}\ \bibnamefont {Liu}}, \bibinfo {author}
  {\bibfnamefont {Z.}~\bibnamefont {Ye}}, \bibinfo {author} {\bibfnamefont
  {Y.}~\bibnamefont {Wang}}, \bibinfo {author} {\bibfnamefont {Y.}~\bibnamefont
  {Alexeev}}, \bibinfo {author} {\bibfnamefont {J.}~\bibnamefont {Eisert}},\
  and\ \bibinfo {author} {\bibfnamefont {L.}~\bibnamefont {Jiang}},\
  }\href@noop {} {\bibinfo {title} {Towards provably efficient quantum
  algorithms for large-scale machine-learning models}} (\bibinfo {year}
  {2023}),\ \Eprint {https://arxiv.org/abs/2303.03428} {arXiv:2303.03428
  [quant-ph]} \BibitemShut {NoStop}%
\bibitem [{\citenamefont {Preskill}(2018)}]{Preskill2018}%
  \BibitemOpen
  \bibfield  {author} {\bibinfo {author} {\bibfnamefont {J.}~\bibnamefont
  {Preskill}},\ }\bibfield  {title} {\bibinfo {title} {Quantum computing in the
  {NISQ} era and beyond},\ }\href {https://doi.org/10.22331/q-2018-08-06-79}
  {\bibfield  {journal} {\bibinfo  {journal} {Quantum}\ }\textbf {\bibinfo
  {volume} {2}},\ \bibinfo {pages} {79} (\bibinfo {year} {2018})}\BibitemShut
  {NoStop}%
\bibitem [{\citenamefont {Liu}\ \emph {et~al.}(2021)\citenamefont {Liu},
  \citenamefont {Arunachalam},\ and\ \citenamefont {Temme}}]{Liu2021}%
  \BibitemOpen
  \bibfield  {author} {\bibinfo {author} {\bibfnamefont {Y.}~\bibnamefont
  {Liu}}, \bibinfo {author} {\bibfnamefont {S.}~\bibnamefont {Arunachalam}},\
  and\ \bibinfo {author} {\bibfnamefont {K.}~\bibnamefont {Temme}},\ }\bibfield
   {title} {\bibinfo {title} {A rigorous and robust quantum speed-up in
  supervised machine learning},\ }\href
  {https://doi.org/10.1038/s41567-021-01287-z} {\bibfield  {journal} {\bibinfo
  {journal} {Nature Physics}\ }\textbf {\bibinfo {volume} {17}},\ \bibinfo
  {pages} {1013} (\bibinfo {year} {2021})}\BibitemShut {NoStop}%
\bibitem [{\citenamefont {Huang}\ \emph {et~al.}(2022)\citenamefont {Huang},
  \citenamefont {Broughton}, \citenamefont {Cotler}, \citenamefont {Chen},
  \citenamefont {Li}, \citenamefont {Mohseni}, \citenamefont {Neven},
  \citenamefont {Babbush}, \citenamefont {Kueng}, \citenamefont {Preskill},\
  and\ \citenamefont {McClean}}]{Huang2022}%
  \BibitemOpen
  \bibfield  {author} {\bibinfo {author} {\bibfnamefont {H.-Y.}\ \bibnamefont
  {Huang}}, \bibinfo {author} {\bibfnamefont {M.}~\bibnamefont {Broughton}},
  \bibinfo {author} {\bibfnamefont {J.}~\bibnamefont {Cotler}}, \bibinfo
  {author} {\bibfnamefont {S.}~\bibnamefont {Chen}}, \bibinfo {author}
  {\bibfnamefont {J.}~\bibnamefont {Li}}, \bibinfo {author} {\bibfnamefont
  {M.}~\bibnamefont {Mohseni}}, \bibinfo {author} {\bibfnamefont
  {H.}~\bibnamefont {Neven}}, \bibinfo {author} {\bibfnamefont
  {R.}~\bibnamefont {Babbush}}, \bibinfo {author} {\bibfnamefont
  {R.}~\bibnamefont {Kueng}}, \bibinfo {author} {\bibfnamefont
  {J.}~\bibnamefont {Preskill}},\ and\ \bibinfo {author} {\bibfnamefont
  {J.~R.}\ \bibnamefont {McClean}},\ }\bibfield  {title} {\bibinfo {title}
  {Quantum advantage in learning from experiments},\ }\href
  {https://doi.org/10.1126/science.abn7293} {\bibfield  {journal} {\bibinfo
  {journal} {Science (New York, N.Y.)}\ }\textbf {\bibinfo {volume} {376}},\
  \bibinfo {pages} {1182} (\bibinfo {year} {2022})}\BibitemShut {NoStop}%
\bibitem [{\citenamefont {{Qiskit Community}}(2017)}]{QiskitCommunity2017}%
  \BibitemOpen
  \bibfield  {author} {\bibinfo {author} {\bibnamefont {{Qiskit Community}}},\
  }\href {https://doi.org/10.5281/zenodo.2562110} {\bibinfo {title} {Qiskit:
  {{An}} open-source framework for quantum computing}} (\bibinfo {year}
  {2017})\BibitemShut {NoStop}%
\bibitem [{\citenamefont {Broughton}\ \emph {et~al.}(2021)\citenamefont
  {Broughton}, \citenamefont {Verdon}, \citenamefont {McCourt}, \citenamefont
  {Martinez}, \citenamefont {Yoo}, \citenamefont {Isakov}, \citenamefont
  {Massey}, \citenamefont {Halavati}, \citenamefont {Niu}, \citenamefont
  {Zlokapa},\ and\ \citenamefont {{others}}}]{broughton2021tensorflow}%
  \BibitemOpen
  \bibfield  {author} {\bibinfo {author} {\bibfnamefont {M.}~\bibnamefont
  {Broughton}}, \bibinfo {author} {\bibfnamefont {G.}~\bibnamefont {Verdon}},
  \bibinfo {author} {\bibfnamefont {T.}~\bibnamefont {McCourt}}, \bibinfo
  {author} {\bibfnamefont {A.~J.}\ \bibnamefont {Martinez}}, \bibinfo {author}
  {\bibfnamefont {J.~H.}\ \bibnamefont {Yoo}}, \bibinfo {author} {\bibfnamefont
  {S.~V.}\ \bibnamefont {Isakov}}, \bibinfo {author} {\bibfnamefont
  {P.}~\bibnamefont {Massey}}, \bibinfo {author} {\bibfnamefont
  {R.}~\bibnamefont {Halavati}}, \bibinfo {author} {\bibfnamefont {M.~Y.}\
  \bibnamefont {Niu}}, \bibinfo {author} {\bibfnamefont {A.}~\bibnamefont
  {Zlokapa}},\ and\ \bibinfo {author} {\bibnamefont {{others}}},\ }\href@noop
  {} {\bibinfo {title} {Tensorflow quantum: A software framework for quantum
  machine learning}} (\bibinfo {year} {2021}),\ \Eprint
  {https://arxiv.org/abs/2003.02989} {arXiv:2003.02989 [quant-ph]} \BibitemShut
  {NoStop}%
\bibitem [{\citenamefont {Bergholm}\ \emph {et~al.}(2022)\citenamefont
  {Bergholm}, \citenamefont {Izaac}, \citenamefont {Schuld}, \citenamefont
  {Gogolin}, \citenamefont {Ahmed}, \citenamefont {Ajith}, \citenamefont
  {Alam}, \citenamefont {Alonso-Linaje}, \citenamefont {AkashNarayanan},
  \citenamefont {Asadi},\ and\ \citenamefont
  {{others}}}]{bergholm2022pennylane}%
  \BibitemOpen
  \bibfield  {author} {\bibinfo {author} {\bibfnamefont {V.}~\bibnamefont
  {Bergholm}}, \bibinfo {author} {\bibfnamefont {J.}~\bibnamefont {Izaac}},
  \bibinfo {author} {\bibfnamefont {M.}~\bibnamefont {Schuld}}, \bibinfo
  {author} {\bibfnamefont {C.}~\bibnamefont {Gogolin}}, \bibinfo {author}
  {\bibfnamefont {S.}~\bibnamefont {Ahmed}}, \bibinfo {author} {\bibfnamefont
  {V.}~\bibnamefont {Ajith}}, \bibinfo {author} {\bibfnamefont {M.~S.}\
  \bibnamefont {Alam}}, \bibinfo {author} {\bibfnamefont {G.}~\bibnamefont
  {Alonso-Linaje}}, \bibinfo {author} {\bibfnamefont {B.}~\bibnamefont
  {AkashNarayanan}}, \bibinfo {author} {\bibfnamefont {A.}~\bibnamefont
  {Asadi}},\ and\ \bibinfo {author} {\bibnamefont {{others}}},\ }\href@noop {}
  {\bibinfo {title} {Pennylane: Automatic differentiation of hybrid
  quantum-classical computations}} (\bibinfo {year} {2022}),\ \Eprint
  {https://arxiv.org/abs/1811.04968} {arXiv:1811.04968 [quant-ph]} \BibitemShut
  {NoStop}%
\bibitem [{\citenamefont {{Cirq Developers}}(2023)}]{cirq2023}%
  \BibitemOpen
  \bibfield  {author} {\bibinfo {author} {\bibnamefont {{Cirq Developers}}},\
  }\href {https://doi.org/10.5281/zenodo.4062499} {\bibinfo {title} {Cirq}}
  (\bibinfo {year} {2023})\BibitemShut {NoStop}%
\bibitem [{\citenamefont {{Di Marcantonio}}\ \emph {et~al.}(2023)\citenamefont
  {{Di Marcantonio}}, \citenamefont {Incudini}, \citenamefont {Tezza},\ and\
  \citenamefont {Grossi}}]{DiMarcantonio2023}%
  \BibitemOpen
  \bibfield  {author} {\bibinfo {author} {\bibfnamefont {F.}~\bibnamefont {{Di
  Marcantonio}}}, \bibinfo {author} {\bibfnamefont {M.}~\bibnamefont
  {Incudini}}, \bibinfo {author} {\bibfnamefont {D.}~\bibnamefont {Tezza}},\
  and\ \bibinfo {author} {\bibfnamefont {M.}~\bibnamefont {Grossi}},\
  }\bibfield  {title} {\bibinfo {title} {Quantum advantage seeker with kernels
  {(QuASK)}: a software framework to speed up the research in quantum machine
  learning},\ }\bibfield  {journal} {\bibinfo  {journal} {Quantum Machine
  Intelligence}\ }\textbf {\bibinfo {volume} {5}},\ \href
  {https://doi.org/10.1007/s42484-023-00107-2} {10.1007/s42484-023-00107-2}
  (\bibinfo {year} {2023})\BibitemShut {NoStop}%
\bibitem [{\citenamefont {Pedregosa}\ \emph {et~al.}(2011)\citenamefont
  {Pedregosa}, \citenamefont {Varoquaux}, \citenamefont {Gramfort},
  \citenamefont {Michel}, \citenamefont {Thirion}, \citenamefont {Grisel},
  \citenamefont {Blondel}, \citenamefont {Prettenhofer}, \citenamefont {Weiss},
  \citenamefont {Dubourg},\ and\ \citenamefont {{others}}}]{Pedregosa2011}%
  \BibitemOpen
  \bibfield  {author} {\bibinfo {author} {\bibfnamefont {F.}~\bibnamefont
  {Pedregosa}}, \bibinfo {author} {\bibfnamefont {G.}~\bibnamefont
  {Varoquaux}}, \bibinfo {author} {\bibfnamefont {A.}~\bibnamefont {Gramfort}},
  \bibinfo {author} {\bibfnamefont {V.}~\bibnamefont {Michel}}, \bibinfo
  {author} {\bibfnamefont {B.}~\bibnamefont {Thirion}}, \bibinfo {author}
  {\bibfnamefont {O.}~\bibnamefont {Grisel}}, \bibinfo {author} {\bibfnamefont
  {M.}~\bibnamefont {Blondel}}, \bibinfo {author} {\bibfnamefont
  {P.}~\bibnamefont {Prettenhofer}}, \bibinfo {author} {\bibfnamefont
  {R.}~\bibnamefont {Weiss}}, \bibinfo {author} {\bibfnamefont
  {V.}~\bibnamefont {Dubourg}},\ and\ \bibinfo {author} {\bibnamefont
  {{others}}},\ }\bibfield  {title} {\bibinfo {title} {Scikit-learn: Machine
  learning in {P}ython},\ }\href@noop {} {\bibfield  {journal} {\bibinfo
  {journal} {Journal of Machine Learning Research}\ }\textbf {\bibinfo {volume}
  {12}},\ \bibinfo {pages} {2825} (\bibinfo {year} {2011})}\BibitemShut
  {NoStop}%
\bibitem [{\citenamefont {Buitinck}\ \emph {et~al.}(2013)\citenamefont
  {Buitinck}, \citenamefont {Louppe}, \citenamefont {Blondel}, \citenamefont
  {Pedregosa}, \citenamefont {Mueller}, \citenamefont {Grisel}, \citenamefont
  {Niculae}, \citenamefont {Prettenhofer}, \citenamefont {Gramfort},
  \citenamefont {Grobler} \emph {et~al.}}]{buitinck2013}%
  \BibitemOpen
  \bibfield  {author} {\bibinfo {author} {\bibfnamefont {L.}~\bibnamefont
  {Buitinck}}, \bibinfo {author} {\bibfnamefont {G.}~\bibnamefont {Louppe}},
  \bibinfo {author} {\bibfnamefont {M.}~\bibnamefont {Blondel}}, \bibinfo
  {author} {\bibfnamefont {F.}~\bibnamefont {Pedregosa}}, \bibinfo {author}
  {\bibfnamefont {A.}~\bibnamefont {Mueller}}, \bibinfo {author} {\bibfnamefont
  {O.}~\bibnamefont {Grisel}}, \bibinfo {author} {\bibfnamefont
  {V.}~\bibnamefont {Niculae}}, \bibinfo {author} {\bibfnamefont
  {P.}~\bibnamefont {Prettenhofer}}, \bibinfo {author} {\bibfnamefont
  {A.}~\bibnamefont {Gramfort}}, \bibinfo {author} {\bibfnamefont
  {J.}~\bibnamefont {Grobler}}, \emph {et~al.},\ }\href@noop {} {\bibinfo
  {title} {Api design for machine learning software: experiences from the
  scikit-learn project}} (\bibinfo {year} {2013}),\ \Eprint
  {https://arxiv.org/abs/1309.0238} {arXiv:1309.0238 [cs.LG]} \BibitemShut
  {NoStop}%
\bibitem [{\citenamefont {Kreuzberger}\ \emph {et~al.}(2022)\citenamefont
  {Kreuzberger}, \citenamefont {K{\"u}hl},\ and\ \citenamefont
  {Hirschl}}]{Kreuzberger2022}%
  \BibitemOpen
  \bibfield  {author} {\bibinfo {author} {\bibfnamefont {D.}~\bibnamefont
  {Kreuzberger}}, \bibinfo {author} {\bibfnamefont {N.}~\bibnamefont
  {K{\"u}hl}},\ and\ \bibinfo {author} {\bibfnamefont {S.}~\bibnamefont
  {Hirschl}},\ }\href@noop {} {\bibinfo {title} {Machine learning operations
  (mlops): Overview, definition, and architecture}} (\bibinfo {year} {2022}),\
  \Eprint {https://arxiv.org/abs/arXiv:2205.02302} {arXiv:arXiv:2205.02302
  [scs.LG]} \BibitemShut {NoStop}%
\bibitem [{\citenamefont {Zaharia}\ \emph {et~al.}(2018)\citenamefont
  {Zaharia}, \citenamefont {Chen}, \citenamefont {Davidson}, \citenamefont
  {Ghodsi}, \citenamefont {Hong}, \citenamefont {Konwinski}, \citenamefont
  {Murching}, \citenamefont {Nykodym}, \citenamefont {Ogilvie}, \citenamefont
  {Parkhe}, \citenamefont {Xie},\ and\ \citenamefont {Zumar}}]{Zaharia2018}%
  \BibitemOpen
  \bibfield  {author} {\bibinfo {author} {\bibfnamefont {M.~A.}\ \bibnamefont
  {Zaharia}}, \bibinfo {author} {\bibfnamefont {A.}~\bibnamefont {Chen}},
  \bibinfo {author} {\bibfnamefont {A.}~\bibnamefont {Davidson}}, \bibinfo
  {author} {\bibfnamefont {A.}~\bibnamefont {Ghodsi}}, \bibinfo {author}
  {\bibfnamefont {S.~A.}\ \bibnamefont {Hong}}, \bibinfo {author}
  {\bibfnamefont {A.}~\bibnamefont {Konwinski}}, \bibinfo {author}
  {\bibfnamefont {S.}~\bibnamefont {Murching}}, \bibinfo {author}
  {\bibfnamefont {T.}~\bibnamefont {Nykodym}}, \bibinfo {author} {\bibfnamefont
  {P.}~\bibnamefont {Ogilvie}}, \bibinfo {author} {\bibfnamefont
  {M.}~\bibnamefont {Parkhe}}, \bibinfo {author} {\bibfnamefont
  {F.}~\bibnamefont {Xie}},\ and\ \bibinfo {author} {\bibfnamefont
  {C.}~\bibnamefont {Zumar}},\ }\bibfield  {title} {\bibinfo {title}
  {{Accelerating the Machine Learning Lifecycle with MLflow}},\ }\href
  {https://api.semanticscholar.org/CorpusID:83459546} {\bibfield  {journal}
  {\bibinfo  {journal} {IEEE Data Eng. Bull.}\ }\textbf {\bibinfo {volume}
  {41}},\ \bibinfo {pages} {39} (\bibinfo {year} {2018})}\BibitemShut {NoStop}%
\bibitem [{\citenamefont {Z{\"o}ller}\ and\ \citenamefont
  {Huber}(2021)}]{Zoeller2021}%
  \BibitemOpen
  \bibfield  {author} {\bibinfo {author} {\bibfnamefont {M.}~\bibnamefont
  {Z{\"o}ller}}\ and\ \bibinfo {author} {\bibfnamefont {M.~F.}\ \bibnamefont
  {Huber}},\ }\href@noop {} {\bibinfo {title} {Benchmark and survey of
  automated machine learning frameworks}} (\bibinfo {year} {2021}),\ \Eprint
  {https://arxiv.org/abs/arXiv:1904.12054v5} {arXiv:arXiv:1904.12054v5 [cs.LG]}
  \BibitemShut {NoStop}%
\bibitem [{\citenamefont {Liaw}\ \emph {et~al.}(2018)\citenamefont {Liaw},
  \citenamefont {Liang}, \citenamefont {Nishihara}, \citenamefont {Moritz},
  \citenamefont {Gonzalez},\ and\ \citenamefont {Stoica}}]{liaw2018tune}%
  \BibitemOpen
  \bibfield  {author} {\bibinfo {author} {\bibfnamefont {R.}~\bibnamefont
  {Liaw}}, \bibinfo {author} {\bibfnamefont {E.}~\bibnamefont {Liang}},
  \bibinfo {author} {\bibfnamefont {R.}~\bibnamefont {Nishihara}}, \bibinfo
  {author} {\bibfnamefont {P.}~\bibnamefont {Moritz}}, \bibinfo {author}
  {\bibfnamefont {J.~E.}\ \bibnamefont {Gonzalez}},\ and\ \bibinfo {author}
  {\bibfnamefont {I.}~\bibnamefont {Stoica}},\ }\bibfield  {title} {\bibinfo
  {title} {Tune: A research platform for distributed model selection and
  training},\ }\href@noop {} {\bibfield  {journal} {\bibinfo  {journal} {arXiv
  preprint arXiv:1807.05118}\ } (\bibinfo {year} {2018})}\BibitemShut {NoStop}%
\bibitem [{\citenamefont {Cerezo}\ \emph {et~al.}(2021)\citenamefont {Cerezo},
  \citenamefont {Arrasmith}, \citenamefont {Babbush}, \citenamefont {Benjamin},
  \citenamefont {Endo}, \citenamefont {Fujii}, \citenamefont {McClean},
  \citenamefont {Mitarai}, \citenamefont {Yuan}, \citenamefont {Cincio},\ and\
  \citenamefont {Coles}}]{Cerezo2021}%
  \BibitemOpen
  \bibfield  {author} {\bibinfo {author} {\bibfnamefont {M.}~\bibnamefont
  {Cerezo}}, \bibinfo {author} {\bibfnamefont {A.}~\bibnamefont {Arrasmith}},
  \bibinfo {author} {\bibfnamefont {R.}~\bibnamefont {Babbush}}, \bibinfo
  {author} {\bibfnamefont {S.~C.}\ \bibnamefont {Benjamin}}, \bibinfo {author}
  {\bibfnamefont {S.}~\bibnamefont {Endo}}, \bibinfo {author} {\bibfnamefont
  {K.}~\bibnamefont {Fujii}}, \bibinfo {author} {\bibfnamefont {J.~R.}\
  \bibnamefont {McClean}}, \bibinfo {author} {\bibfnamefont {K.}~\bibnamefont
  {Mitarai}}, \bibinfo {author} {\bibfnamefont {X.}~\bibnamefont {Yuan}},
  \bibinfo {author} {\bibfnamefont {L.}~\bibnamefont {Cincio}},\ and\ \bibinfo
  {author} {\bibfnamefont {P.~J.}\ \bibnamefont {Coles}},\ }\bibfield  {title}
  {\bibinfo {title} {Variational quantum algorithms},\ }\bibfield  {journal}
  {\bibinfo  {journal} {Nature Reviews Physics}\ }\href
  {https://doi.org/10.1038/s42254-021-00348-9} {10.1038/s42254-021-00348-9}
  (\bibinfo {year} {2021})\BibitemShut {NoStop}%
\bibitem [{\citenamefont {Cong}\ \emph {et~al.}(2019)\citenamefont {Cong},
  \citenamefont {Choi},\ and\ \citenamefont {Lukin}}]{Cong_2019}%
  \BibitemOpen
  \bibfield  {author} {\bibinfo {author} {\bibfnamefont {I.}~\bibnamefont
  {Cong}}, \bibinfo {author} {\bibfnamefont {S.}~\bibnamefont {Choi}},\ and\
  \bibinfo {author} {\bibfnamefont {M.~D.}\ \bibnamefont {Lukin}},\ }\bibfield
  {title} {\bibinfo {title} {Quantum convolutional neural networks},\ }\href
  {https://doi.org/10.1038/s41567-019-0648-8} {\bibfield  {journal} {\bibinfo
  {journal} {Nature Physics}\ }\textbf {\bibinfo {volume} {15}},\ \bibinfo
  {pages} {1273–1278} (\bibinfo {year} {2019})}\BibitemShut {NoStop}%
\bibitem [{\citenamefont {Schuld}\ and\ \citenamefont
  {Killoran}(2019)}]{Schuld2019}%
  \BibitemOpen
  \bibfield  {author} {\bibinfo {author} {\bibfnamefont {M.}~\bibnamefont
  {Schuld}}\ and\ \bibinfo {author} {\bibfnamefont {N.}~\bibnamefont
  {Killoran}},\ }\bibfield  {title} {\bibinfo {title} {Quantum machine learning
  in feature hilbert spaces},\ }\href
  {https://doi.org/10.1103/PhysRevLett.122.040504} {\bibfield  {journal}
  {\bibinfo  {journal} {Phys. Rev. Lett.}\ }\textbf {\bibinfo {volume} {122}},\
  \bibinfo {pages} {040504} (\bibinfo {year} {2019})}\BibitemShut {NoStop}%
\bibitem [{\citenamefont {Havl{\'i}{\v{c}}ek}\ \emph
  {et~al.}(2019)\citenamefont {Havl{\'i}{\v{c}}ek}, \citenamefont
  {C{\'o}rcoles}, \citenamefont {Temme}, \citenamefont {Harrow}, \citenamefont
  {Kandala}, \citenamefont {Chow},\ and\ \citenamefont
  {Gambetta}}]{Havlicek2019}%
  \BibitemOpen
  \bibfield  {author} {\bibinfo {author} {\bibfnamefont {V.}~\bibnamefont
  {Havl{\'i}{\v{c}}ek}}, \bibinfo {author} {\bibfnamefont {A.~D.}\ \bibnamefont
  {C{\'o}rcoles}}, \bibinfo {author} {\bibfnamefont {K.}~\bibnamefont {Temme}},
  \bibinfo {author} {\bibfnamefont {A.~W.}\ \bibnamefont {Harrow}}, \bibinfo
  {author} {\bibfnamefont {A.}~\bibnamefont {Kandala}}, \bibinfo {author}
  {\bibfnamefont {J.~M.}\ \bibnamefont {Chow}},\ and\ \bibinfo {author}
  {\bibfnamefont {J.~M.}\ \bibnamefont {Gambetta}},\ }\bibfield  {title}
  {\bibinfo {title} {Supervised learning with quantum-enhanced feature
  spaces},\ }\href {https://doi.org/10.1038/s41586-019-0980-2} {\bibfield
  {journal} {\bibinfo  {journal} {Nature}\ }\textbf {\bibinfo {volume} {567}},\
  \bibinfo {pages} {209} (\bibinfo {year} {2019})}\BibitemShut {NoStop}%
\bibitem [{\citenamefont {Rapp}\ and\ \citenamefont {Roth}(2023)}]{Rapp2023}%
  \BibitemOpen
  \bibfield  {author} {\bibinfo {author} {\bibfnamefont {F.}~\bibnamefont
  {Rapp}}\ and\ \bibinfo {author} {\bibfnamefont {M.}~\bibnamefont {Roth}},\
  }\bibfield  {title} {\bibinfo {title} {Quantum gaussian process regression
  for bayesian optimization},\ }\href {http://arxiv.org/pdf/2304.12923v1}
  {\bibfield  {journal} {\bibinfo  {journal} {arXiv preprint arXiv:2304.12923}\
  } (\bibinfo {year} {2023})}\BibitemShut {NoStop}%
\bibitem [{\citenamefont {Haug}\ \emph {et~al.}(2021)\citenamefont {Haug},
  \citenamefont {Bharti},\ and\ \citenamefont {Kim}}]{Haug2021}%
  \BibitemOpen
  \bibfield  {author} {\bibinfo {author} {\bibfnamefont {T.}~\bibnamefont
  {Haug}}, \bibinfo {author} {\bibfnamefont {K.}~\bibnamefont {Bharti}},\ and\
  \bibinfo {author} {\bibfnamefont {M.}~\bibnamefont {Kim}},\ }\bibfield
  {title} {\bibinfo {title} {Capacity and quantum geometry of parametrized
  quantum circuits},\ }\href {https://doi.org/10.1103/PRXQuantum.2.040309}
  {\bibfield  {journal} {\bibinfo  {journal} {PRX Quantum}\ }\textbf {\bibinfo
  {volume} {2}},\ \bibinfo {pages} {040309} (\bibinfo {year}
  {2021})}\BibitemShut {NoStop}%
\bibitem [{\citenamefont {Huang}\ \emph {et~al.}(2021)\citenamefont {Huang},
  \citenamefont {Broughton}, \citenamefont {Mohseni}, \citenamefont {Babbush},
  \citenamefont {Boixo}, \citenamefont {Neven},\ and\ \citenamefont
  {Mcclean}}]{Huang2021}%
  \BibitemOpen
  \bibfield  {author} {\bibinfo {author} {\bibfnamefont {H.-Y.}\ \bibnamefont
  {Huang}}, \bibinfo {author} {\bibfnamefont {M.}~\bibnamefont {Broughton}},
  \bibinfo {author} {\bibfnamefont {M.}~\bibnamefont {Mohseni}}, \bibinfo
  {author} {\bibfnamefont {R.}~\bibnamefont {Babbush}}, \bibinfo {author}
  {\bibfnamefont {S.}~\bibnamefont {Boixo}}, \bibinfo {author} {\bibfnamefont
  {H.}~\bibnamefont {Neven}},\ and\ \bibinfo {author} {\bibfnamefont
  {J.}~\bibnamefont {Mcclean}},\ }\bibfield  {title} {\bibinfo {title} {Power
  of data in quantum machine learning},\ }\href
  {https://doi.org/10.1038/s41467-021-22539-9} {\bibfield  {journal} {\bibinfo
  {journal} {Nature Communications}\ }\textbf {\bibinfo {volume} {12}}
  (\bibinfo {year} {2021})}\BibitemShut {NoStop}%
\bibitem [{\citenamefont {Abadi}\ \emph {et~al.}(2015)\citenamefont {Abadi},
  \citenamefont {Agarwal}, \citenamefont {Barham}, \citenamefont {Brevdo},
  \citenamefont {Chen}, \citenamefont {Citro}, \citenamefont {Corrado},
  \citenamefont {Davis}, \citenamefont {Dean}, \citenamefont {Devin},\ and\
  \citenamefont {{others}}}]{tensorflow2015}%
  \BibitemOpen
  \bibfield  {author} {\bibinfo {author} {\bibfnamefont {M.}~\bibnamefont
  {Abadi}}, \bibinfo {author} {\bibfnamefont {A.}~\bibnamefont {Agarwal}},
  \bibinfo {author} {\bibfnamefont {P.}~\bibnamefont {Barham}}, \bibinfo
  {author} {\bibfnamefont {E.}~\bibnamefont {Brevdo}}, \bibinfo {author}
  {\bibfnamefont {Z.}~\bibnamefont {Chen}}, \bibinfo {author} {\bibfnamefont
  {C.}~\bibnamefont {Citro}}, \bibinfo {author} {\bibfnamefont {G.~S.}\
  \bibnamefont {Corrado}}, \bibinfo {author} {\bibfnamefont {A.}~\bibnamefont
  {Davis}}, \bibinfo {author} {\bibfnamefont {J.}~\bibnamefont {Dean}},
  \bibinfo {author} {\bibfnamefont {M.}~\bibnamefont {Devin}},\ and\ \bibinfo
  {author} {\bibnamefont {{others}}},\ }\href {https://www.tensorflow.org/}
  {\bibinfo {title} {{TensorFlow}: Large-scale machine learning on
  heterogeneous systems}} (\bibinfo {year} {2015}),\ \bibinfo {note} {software
  available from tensorflow.org}\BibitemShut {NoStop}%
\bibitem [{\citenamefont {Chollet}\ \emph {et~al.}(2015)\citenamefont {Chollet}
  \emph {et~al.}}]{Chollet2015}%
  \BibitemOpen
  \bibfield  {author} {\bibinfo {author} {\bibfnamefont {F.}~\bibnamefont
  {Chollet}} \emph {et~al.},\ }\href@noop {} {\bibinfo {title} {Keras}},\
  \bibinfo {howpublished} {\url{https://keras.io}} (\bibinfo {year}
  {2015})\BibitemShut {NoStop}%
\bibitem [{\citenamefont {Paszke}\ \emph {et~al.}(2019)\citenamefont {Paszke},
  \citenamefont {Gross}, \citenamefont {Massa}, \citenamefont {Lerer},
  \citenamefont {Bradbury}, \citenamefont {Chanan}, \citenamefont {Killeen},
  \citenamefont {Lin}, \citenamefont {Gimelshein}, \citenamefont {Antiga},\
  and\ \citenamefont {{others}}}]{Paszke2019}%
  \BibitemOpen
  \bibfield  {author} {\bibinfo {author} {\bibfnamefont {A.}~\bibnamefont
  {Paszke}}, \bibinfo {author} {\bibfnamefont {S.}~\bibnamefont {Gross}},
  \bibinfo {author} {\bibfnamefont {F.}~\bibnamefont {Massa}}, \bibinfo
  {author} {\bibfnamefont {A.}~\bibnamefont {Lerer}}, \bibinfo {author}
  {\bibfnamefont {J.}~\bibnamefont {Bradbury}}, \bibinfo {author}
  {\bibfnamefont {G.}~\bibnamefont {Chanan}}, \bibinfo {author} {\bibfnamefont
  {T.}~\bibnamefont {Killeen}}, \bibinfo {author} {\bibfnamefont
  {Z.}~\bibnamefont {Lin}}, \bibinfo {author} {\bibfnamefont {N.}~\bibnamefont
  {Gimelshein}}, \bibinfo {author} {\bibfnamefont {L.}~\bibnamefont {Antiga}},\
  and\ \bibinfo {author} {\bibnamefont {{others}}},\ }\bibfield  {title}
  {\bibinfo {title} {{PyTorch: An Imperative Style, High-Performance Deep
  Learning Library}},\ }in\ \href
  {http://papers.neurips.cc/paper/9015-pytorch-an-imperative-style-high-performance-deep-learning-library.pdf}
  {\emph {\bibinfo {booktitle} {Advances in Neural Information Processing
  Systems 32}}},\ \bibinfo {editor} {edited by\ \bibinfo {editor}
  {\bibfnamefont {H.}~\bibnamefont {Wallach}}, \bibinfo {editor} {\bibfnamefont
  {H.}~\bibnamefont {Larochelle}}, \bibinfo {editor} {\bibfnamefont
  {A.}~\bibnamefont {Beygelzimer}}, \bibinfo {editor} {\bibfnamefont
  {F.}~\bibnamefont {d'Alché Buc}}, \bibinfo {editor} {\bibfnamefont
  {E.}~\bibnamefont {Fox}},\ and\ \bibinfo {editor} {\bibfnamefont
  {R.}~\bibnamefont {Garnett}}}\ (\bibinfo  {publisher} {Curran Associates,
  Inc.},\ \bibinfo {year} {2019})\ pp.\ \bibinfo {pages}
  {8024--8035}\BibitemShut {NoStop}%
\bibitem [{\citenamefont {Schuld}\ and\ \citenamefont
  {Petruccione}(2018)}]{Schuld2018}%
  \BibitemOpen
  \bibfield  {author} {\bibinfo {author} {\bibfnamefont {M.}~\bibnamefont
  {Schuld}}\ and\ \bibinfo {author} {\bibfnamefont {F.}~\bibnamefont
  {Petruccione}},\ }\href@noop {} {\emph {\bibinfo {title} {Supervised learning
  with quantum computers}}},\ Vol.~\bibinfo {volume} {17}\ (\bibinfo
  {publisher} {Springer},\ \bibinfo {year} {2018})\BibitemShut {NoStop}%
\bibitem [{\citenamefont {Liu}\ \emph {et~al.}(2019)\citenamefont {Liu},
  \citenamefont {Yuan}, \citenamefont {Lu},\ and\ \citenamefont
  {Wang}}]{Liu2020}%
  \BibitemOpen
  \bibfield  {author} {\bibinfo {author} {\bibfnamefont {J.}~\bibnamefont
  {Liu}}, \bibinfo {author} {\bibfnamefont {H.}~\bibnamefont {Yuan}}, \bibinfo
  {author} {\bibfnamefont {X.-M.}\ \bibnamefont {Lu}},\ and\ \bibinfo {author}
  {\bibfnamefont {X.}~\bibnamefont {Wang}},\ }\bibfield  {title} {\bibinfo
  {title} {Quantum fisher information matrix and multiparameter estimation},\
  }\href {https://doi.org/10.1088/1751-8121/ab5d4d} {\bibfield  {journal}
  {\bibinfo  {journal} {Journal of Physics A: Mathematical and Theoretical}\
  }\textbf {\bibinfo {volume} {53}},\ \bibinfo {pages} {023001} (\bibinfo
  {year} {2019})}\BibitemShut {NoStop}%
\bibitem [{\citenamefont {Meyer}(2021)}]{Meyer2021}%
  \BibitemOpen
  \bibfield  {author} {\bibinfo {author} {\bibfnamefont {J.~J.}\ \bibnamefont
  {Meyer}},\ }\bibfield  {title} {\bibinfo {title} {Fisher {I}nformation in
  {N}oisy {I}ntermediate-{S}cale {Q}uantum {A}pplications},\ }\href
  {https://doi.org/10.22331/q-2021-09-09-539} {\bibfield  {journal} {\bibinfo
  {journal} {{Quantum}}\ }\textbf {\bibinfo {volume} {5}},\ \bibinfo {pages}
  {539} (\bibinfo {year} {2021})}\BibitemShut {NoStop}%
\bibitem [{\citenamefont {Peters}\ \emph {et~al.}(2021)\citenamefont {Peters},
  \citenamefont {Caldeira}, \citenamefont {Ho}, \citenamefont {Leichenauer},
  \citenamefont {Mohseni}, \citenamefont {Neven}, \citenamefont {Spentzouris},
  \citenamefont {Strain},\ and\ \citenamefont {Perdue}}]{Peters2021}%
  \BibitemOpen
  \bibfield  {author} {\bibinfo {author} {\bibfnamefont {E.}~\bibnamefont
  {Peters}}, \bibinfo {author} {\bibfnamefont {J.}~\bibnamefont {Caldeira}},
  \bibinfo {author} {\bibfnamefont {A.}~\bibnamefont {Ho}}, \bibinfo {author}
  {\bibfnamefont {S.}~\bibnamefont {Leichenauer}}, \bibinfo {author}
  {\bibfnamefont {M.}~\bibnamefont {Mohseni}}, \bibinfo {author} {\bibfnamefont
  {H.}~\bibnamefont {Neven}}, \bibinfo {author} {\bibfnamefont
  {P.}~\bibnamefont {Spentzouris}}, \bibinfo {author} {\bibfnamefont
  {D.}~\bibnamefont {Strain}},\ and\ \bibinfo {author} {\bibfnamefont {G.~N.}\
  \bibnamefont {Perdue}},\ }\bibfield  {title} {\bibinfo {title} {Machine
  learning of high dimensional data on a noisy quantum processor},\ }\href
  {https://doi.org/10.1038/s41534-021-00498-9} {\bibfield  {journal} {\bibinfo
  {journal} {npj Quantum Information}\ }\textbf {\bibinfo {volume} {7}},\
  \bibinfo {pages} {161} (\bibinfo {year} {2021})}\BibitemShut {NoStop}%
\bibitem [{\citenamefont {Schuld}(2021)}]{schuld2021supervised}%
  \BibitemOpen
  \bibfield  {author} {\bibinfo {author} {\bibfnamefont {M.}~\bibnamefont
  {Schuld}},\ }\href@noop {} {\bibinfo {title} {Supervised quantum machine
  learning models are kernel methods}} (\bibinfo {year} {2021}),\ \Eprint
  {https://arxiv.org/abs/2101.11020} {arXiv:2101.11020 [quant-ph]} \BibitemShut
  {NoStop}%
\bibitem [{\citenamefont {Vapnik}(1999)}]{vapnik1999}%
  \BibitemOpen
  \bibfield  {author} {\bibinfo {author} {\bibfnamefont {V.}~\bibnamefont
  {Vapnik}},\ }\href@noop {} {\emph {\bibinfo {title} {The nature of
  statistical learning theory}}}\ (\bibinfo  {publisher} {Springer science \&
  business media},\ \bibinfo {year} {1999})\BibitemShut {NoStop}%
\bibitem [{\citenamefont {Murphy}(2012)}]{Murphy2012}%
  \BibitemOpen
  \bibfield  {author} {\bibinfo {author} {\bibfnamefont {K.~P.}\ \bibnamefont
  {Murphy}},\ }\href@noop {} {\emph {\bibinfo {title} {Machine learning: a
  probabilistic perspective}}}\ (\bibinfo  {publisher} {MIT press},\ \bibinfo
  {year} {2012})\BibitemShut {NoStop}%
\bibitem [{\citenamefont {Mangini}\ \emph {et~al.}(2021)\citenamefont
  {Mangini}, \citenamefont {Tacchino}, \citenamefont {Gerace}, \citenamefont
  {Bajoni},\ and\ \citenamefont {Macchiavello}}]{Mangini2021}%
  \BibitemOpen
  \bibfield  {author} {\bibinfo {author} {\bibfnamefont {S.}~\bibnamefont
  {Mangini}}, \bibinfo {author} {\bibfnamefont {F.}~\bibnamefont {Tacchino}},
  \bibinfo {author} {\bibfnamefont {D.}~\bibnamefont {Gerace}}, \bibinfo
  {author} {\bibfnamefont {D.}~\bibnamefont {Bajoni}},\ and\ \bibinfo {author}
  {\bibfnamefont {C.}~\bibnamefont {Macchiavello}},\ }\bibfield  {title}
  {\bibinfo {title} {Quantum computing models for artificial neural networks},\
  }\href {https://doi.org/10.1209/0295-5075/134/10002} {\bibfield  {journal}
  {\bibinfo  {journal} {Europhysics Letters}\ }\textbf {\bibinfo {volume}
  {134}},\ \bibinfo {pages} {10002} (\bibinfo {year} {2021})}\BibitemShut
  {NoStop}%
\bibitem [{\citenamefont {Harrow}\ and\ \citenamefont
  {Napp}(2021)}]{Harrow2021}%
  \BibitemOpen
  \bibfield  {author} {\bibinfo {author} {\bibfnamefont {A.~W.}\ \bibnamefont
  {Harrow}}\ and\ \bibinfo {author} {\bibfnamefont {J.~C.}\ \bibnamefont
  {Napp}},\ }\bibfield  {title} {\bibinfo {title} {Low-depth gradient
  measurements can improve convergence in variational hybrid quantum-classical
  algorithms},\ }\href {https://doi.org/10.1103/PhysRevLett.126.140502}
  {\bibfield  {journal} {\bibinfo  {journal} {Phys. Rev. Lett.}\ }\textbf
  {\bibinfo {volume} {126}},\ \bibinfo {pages} {140502} (\bibinfo {year}
  {2021})}\BibitemShut {NoStop}%
\bibitem [{\citenamefont {Mitarai}\ \emph {et~al.}(2018)\citenamefont
  {Mitarai}, \citenamefont {Negoro}, \citenamefont {Kitagawa},\ and\
  \citenamefont {Fujii}}]{Mitarai2018}%
  \BibitemOpen
  \bibfield  {author} {\bibinfo {author} {\bibfnamefont {K.}~\bibnamefont
  {Mitarai}}, \bibinfo {author} {\bibfnamefont {M.}~\bibnamefont {Negoro}},
  \bibinfo {author} {\bibfnamefont {M.}~\bibnamefont {Kitagawa}},\ and\
  \bibinfo {author} {\bibfnamefont {K.}~\bibnamefont {Fujii}},\ }\bibfield
  {title} {\bibinfo {title} {Quantum circuit learning},\ }\href
  {https://doi.org/10.1103/PhysRevA.98.032309} {\bibfield  {journal} {\bibinfo
  {journal} {Phys. Rev. A}\ }\textbf {\bibinfo {volume} {98}},\ \bibinfo
  {pages} {032309} (\bibinfo {year} {2018})}\BibitemShut {NoStop}%
\bibitem [{\citenamefont {Kyriienko}\ \emph {et~al.}(2021)\citenamefont
  {Kyriienko}, \citenamefont {Paine},\ and\ \citenamefont
  {Elfving}}]{Kyriienko2021}%
  \BibitemOpen
  \bibfield  {author} {\bibinfo {author} {\bibfnamefont {O.}~\bibnamefont
  {Kyriienko}}, \bibinfo {author} {\bibfnamefont {A.~E.}\ \bibnamefont
  {Paine}},\ and\ \bibinfo {author} {\bibfnamefont {V.~E.}\ \bibnamefont
  {Elfving}},\ }\bibfield  {title} {\bibinfo {title} {Solving nonlinear
  differential equations with differentiable quantum circuits},\ }\href
  {https://doi.org/10.1103/PhysRevA.103.052416} {\bibfield  {journal} {\bibinfo
   {journal} {Phys. Rev. A}\ }\textbf {\bibinfo {volume} {103}},\ \bibinfo
  {pages} {052416} (\bibinfo {year} {2021})}\BibitemShut {NoStop}%
\bibitem [{\citenamefont {Gujju}\ \emph {et~al.}(2023)\citenamefont {Gujju},
  \citenamefont {Matsuo},\ and\ \citenamefont {Raymond}}]{Gujju2023}%
  \BibitemOpen
  \bibfield  {author} {\bibinfo {author} {\bibfnamefont {Y.}~\bibnamefont
  {Gujju}}, \bibinfo {author} {\bibfnamefont {A.}~\bibnamefont {Matsuo}},\ and\
  \bibinfo {author} {\bibfnamefont {R.}~\bibnamefont {Raymond}},\ }\href@noop
  {} {\bibinfo {title} {Quantum machine learning on near-term quantum devices:
  Current state of supervised and unsupervised techniques for real-world
  applications}} (\bibinfo {year} {2023}),\ \Eprint
  {https://arxiv.org/abs/2307.00908} {arXiv:2307.00908 [quant-ph]} \BibitemShut
  {NoStop}%
\bibitem [{\citenamefont {Schuld}\ \emph {et~al.}(2021)\citenamefont {Schuld},
  \citenamefont {Sweke},\ and\ \citenamefont {Meyer}}]{Schuld2021}%
  \BibitemOpen
  \bibfield  {author} {\bibinfo {author} {\bibfnamefont {M.}~\bibnamefont
  {Schuld}}, \bibinfo {author} {\bibfnamefont {R.}~\bibnamefont {Sweke}},\ and\
  \bibinfo {author} {\bibfnamefont {J.~J.}\ \bibnamefont {Meyer}},\ }\bibfield
  {title} {\bibinfo {title} {The effect of data encoding on the expressive
  power of variational quantum-machine-learning models},\ }\href
  {https://doi.org/10.1103/PhysRevA.103.032430} {\bibfield  {journal} {\bibinfo
   {journal} {Phys. Rev. A}\ }\textbf {\bibinfo {volume} {103}},\ \bibinfo
  {pages} {032430} (\bibinfo {year} {2021})}\BibitemShut {NoStop}%
\bibitem [{\citenamefont {Caro}\ \emph {et~al.}(2022)\citenamefont {Caro},
  \citenamefont {Huang}, \citenamefont {Cerezo}, \citenamefont {Sharma},
  \citenamefont {Sornborger}, \citenamefont {Cincio},\ and\ \citenamefont
  {Coles}}]{Caro2022}%
  \BibitemOpen
  \bibfield  {author} {\bibinfo {author} {\bibfnamefont {M.~C.}\ \bibnamefont
  {Caro}}, \bibinfo {author} {\bibfnamefont {H.-Y.}\ \bibnamefont {Huang}},
  \bibinfo {author} {\bibfnamefont {M.}~\bibnamefont {Cerezo}}, \bibinfo
  {author} {\bibfnamefont {K.}~\bibnamefont {Sharma}}, \bibinfo {author}
  {\bibfnamefont {A.}~\bibnamefont {Sornborger}}, \bibinfo {author}
  {\bibfnamefont {L.}~\bibnamefont {Cincio}},\ and\ \bibinfo {author}
  {\bibfnamefont {P.~J.}\ \bibnamefont {Coles}},\ }\bibfield  {title} {\bibinfo
  {title} {Generalization in quantum machine learning from few training data},\
  }\href {https://doi.org/10.1038/s41467-022-32550-3} {\bibfield  {journal}
  {\bibinfo  {journal} {Nature communications}\ }\textbf {\bibinfo {volume}
  {13}},\ \bibinfo {pages} {4919} (\bibinfo {year} {2022})}\BibitemShut
  {NoStop}%
\bibitem [{\citenamefont {Kreplin}\ and\ \citenamefont
  {Roth}(2023)}]{Kreplin2023}%
  \BibitemOpen
  \bibfield  {author} {\bibinfo {author} {\bibfnamefont {D.~A.}\ \bibnamefont
  {Kreplin}}\ and\ \bibinfo {author} {\bibfnamefont {M.}~\bibnamefont {Roth}},\
  }\href@noop {} {\bibinfo {title} {Reduction of finite sampling noise in
  quantum neural networks}} (\bibinfo {year} {2023}),\ \Eprint
  {https://arxiv.org/abs/2306.01639} {arXiv:2306.01639 [quant-ph]} \BibitemShut
  {NoStop}%
\bibitem [{squ(2023)}]{squlearnDocu2023}%
  \BibitemOpen
  \href@noop {} {\bibinfo {title} {{sQUlearn Documentation -- 0.5.0}}},\
  \bibinfo {howpublished} {\url{https://squlearn.github.io}} (\bibinfo {year}
  {2023}),\ \bibinfo {note} {accessed: 2023-11-14}\BibitemShut {NoStop}%
\bibitem [{\citenamefont {Pesah}\ \emph {et~al.}(2021)\citenamefont {Pesah},
  \citenamefont {Cerezo}, \citenamefont {Wang}, \citenamefont {Volkoff},
  \citenamefont {Sornborger},\ and\ \citenamefont {Coles}}]{Pesah2021}%
  \BibitemOpen
  \bibfield  {author} {\bibinfo {author} {\bibfnamefont {A.}~\bibnamefont
  {Pesah}}, \bibinfo {author} {\bibfnamefont {M.}~\bibnamefont {Cerezo}},
  \bibinfo {author} {\bibfnamefont {S.}~\bibnamefont {Wang}}, \bibinfo {author}
  {\bibfnamefont {T.}~\bibnamefont {Volkoff}}, \bibinfo {author} {\bibfnamefont
  {A.~T.}\ \bibnamefont {Sornborger}},\ and\ \bibinfo {author} {\bibfnamefont
  {P.~J.}\ \bibnamefont {Coles}},\ }\bibfield  {title} {\bibinfo {title}
  {Absence of barren plateaus in quantum convolutional neural networks},\
  }\href {https://doi.org/10.1103/PhysRevX.11.041011} {\bibfield  {journal}
  {\bibinfo  {journal} {Phys. Rev. X}\ }\textbf {\bibinfo {volume} {11}},\
  \bibinfo {pages} {041011} (\bibinfo {year} {2021})}\BibitemShut {NoStop}%
\bibitem [{\citenamefont {Haug}\ \emph {et~al.}(2023)\citenamefont {Haug},
  \citenamefont {Self},\ and\ \citenamefont {Kim}}]{Haug2023}%
  \BibitemOpen
  \bibfield  {author} {\bibinfo {author} {\bibfnamefont {T.}~\bibnamefont
  {Haug}}, \bibinfo {author} {\bibfnamefont {C.~N.}\ \bibnamefont {Self}},\
  and\ \bibinfo {author} {\bibfnamefont {M.~S.}\ \bibnamefont {Kim}},\
  }\bibfield  {title} {\bibinfo {title} {Quantum machine learning of large
  datasets using randomized measurements},\ }\href
  {https://doi.org/10.1088/2632-2153/acb0b4} {\bibfield  {journal} {\bibinfo
  {journal} {Machine Learning: Science and Technology}\ }\textbf {\bibinfo
  {volume} {4}},\ \bibinfo {pages} {015005} (\bibinfo {year}
  {2023})}\BibitemShut {NoStop}%
\bibitem [{\citenamefont {McClean}\ \emph {et~al.}(2016)\citenamefont
  {McClean}, \citenamefont {Romero}, \citenamefont {Babbush},\ and\
  \citenamefont {Aspuru-Guzik}}]{McClean2016}%
  \BibitemOpen
  \bibfield  {author} {\bibinfo {author} {\bibfnamefont {J.~R.}\ \bibnamefont
  {McClean}}, \bibinfo {author} {\bibfnamefont {J.}~\bibnamefont {Romero}},
  \bibinfo {author} {\bibfnamefont {R.}~\bibnamefont {Babbush}},\ and\ \bibinfo
  {author} {\bibfnamefont {A.}~\bibnamefont {Aspuru-Guzik}},\ }\bibfield
  {title} {\bibinfo {title} {The theory of variational hybrid quantum-classical
  algorithms},\ }\href {https://doi.org/10.1088/1367-2630/18/2/023023}
  {\bibfield  {journal} {\bibinfo  {journal} {New Journal of Physics}\ }\textbf
  {\bibinfo {volume} {18}},\ \bibinfo {pages} {023023} (\bibinfo {year}
  {2016})}\BibitemShut {NoStop}%
\bibitem [{\citenamefont {Sch{\"o}lkopf}\ and\ \citenamefont
  {Smola}(2002)}]{schölkopf2002learning}%
  \BibitemOpen
  \bibfield  {author} {\bibinfo {author} {\bibfnamefont {B.}~\bibnamefont
  {Sch{\"o}lkopf}}\ and\ \bibinfo {author} {\bibfnamefont {A.}~\bibnamefont
  {Smola}},\ }\href {https://books.google.de/books?id=y8ORL3DWt4sC} {\emph
  {\bibinfo {title} {Learning with Kernels: Support Vector Machines,
  Regularization, Optimization, and Beyond}}},\ Adaptive computation and
  machine learning\ (\bibinfo  {publisher} {MIT Press},\ \bibinfo {year}
  {2002})\BibitemShut {NoStop}%
\bibitem [{\citenamefont {Jerbi}\ \emph {et~al.}(2023)\citenamefont {Jerbi},
  \citenamefont {Fiderer}, \citenamefont {Poulsen~Nautrup}, \citenamefont
  {Kübler}, \citenamefont {Briegel},\ and\ \citenamefont
  {Dunjko}}]{Jerbi2023}%
  \BibitemOpen
  \bibfield  {author} {\bibinfo {author} {\bibfnamefont {S.}~\bibnamefont
  {Jerbi}}, \bibinfo {author} {\bibfnamefont {L.~J.}\ \bibnamefont {Fiderer}},
  \bibinfo {author} {\bibfnamefont {H.}~\bibnamefont {Poulsen~Nautrup}},
  \bibinfo {author} {\bibfnamefont {J.~M.}\ \bibnamefont {Kübler}}, \bibinfo
  {author} {\bibfnamefont {H.~J.}\ \bibnamefont {Briegel}},\ and\ \bibinfo
  {author} {\bibfnamefont {V.}~\bibnamefont {Dunjko}},\ }\bibfield  {title}
  {\bibinfo {title} {Quantum machine learning beyond kernel methods},\ }\href
  {https://doi.org/10.1038/s41467-023-36159-y} {\bibfield  {journal} {\bibinfo
  {journal} {Nat. Commun.}\ }\textbf {\bibinfo {volume} {14}},\ \bibinfo
  {pages} {517} (\bibinfo {year} {2023})}\BibitemShut {NoStop}%
\bibitem [{\citenamefont {Thanasilp}\ \emph {et~al.}(2022)\citenamefont
  {Thanasilp}, \citenamefont {Wang}, \citenamefont {Cerezo},\ and\
  \citenamefont {Holmes}}]{thanasilp2022}%
  \BibitemOpen
  \bibfield  {author} {\bibinfo {author} {\bibfnamefont {S.}~\bibnamefont
  {Thanasilp}}, \bibinfo {author} {\bibfnamefont {S.}~\bibnamefont {Wang}},
  \bibinfo {author} {\bibfnamefont {M.}~\bibnamefont {Cerezo}},\ and\ \bibinfo
  {author} {\bibfnamefont {Z.}~\bibnamefont {Holmes}},\ }\href@noop {}
  {\bibinfo {title} {Exponential concentration and untrainability in quantum
  kernel methods}} (\bibinfo {year} {2022}),\ \Eprint
  {https://arxiv.org/abs/2208.11060} {arXiv:2208.11060 [quant-ph]} \BibitemShut
  {NoStop}%
\bibitem [{\citenamefont {Kandola}\ and\ \citenamefont
  {Shawe-Taylor}(2003)}]{Kandola2003}%
  \BibitemOpen
  \bibfield  {author} {\bibinfo {author} {\bibfnamefont {J.~S.}\ \bibnamefont
  {Kandola}}\ and\ \bibinfo {author} {\bibfnamefont {J.}~\bibnamefont
  {Shawe-Taylor}},\ }\bibfield  {title} {\bibinfo {title} {Refining kernels for
  regression and uneven classification problems},\ }in\ \href
  {https://proceedings.mlr.press/r4/kandola03a.html} {\emph {\bibinfo
  {booktitle} {Proceedings of the Ninth International Workshop on Artificial
  Intelligence and Statistics}}},\ \bibinfo {series} {Proceedings of Machine
  Learning Research}, Vol.~\bibinfo {volume} {R4},\ \bibinfo {editor} {edited
  by\ \bibinfo {editor} {\bibfnamefont {C.~M.}\ \bibnamefont {Bishop}}\ and\
  \bibinfo {editor} {\bibfnamefont {B.~J.}\ \bibnamefont {Frey}}}\ (\bibinfo
  {publisher} {PMLR},\ \bibinfo {year} {2003})\ pp.\ \bibinfo {pages}
  {157--162}\BibitemShut {NoStop}%
\bibitem [{\citenamefont {Cristianini}\ \emph {et~al.}(2006)\citenamefont
  {Cristianini}, \citenamefont {Kandola}, \citenamefont {Elisseeff},\ and\
  \citenamefont {Shawe-Taylor}}]{Cristianini2006}%
  \BibitemOpen
  \bibfield  {author} {\bibinfo {author} {\bibfnamefont {N.}~\bibnamefont
  {Cristianini}}, \bibinfo {author} {\bibfnamefont {J.}~\bibnamefont
  {Kandola}}, \bibinfo {author} {\bibfnamefont {A.}~\bibnamefont {Elisseeff}},\
  and\ \bibinfo {author} {\bibfnamefont {J.}~\bibnamefont {Shawe-Taylor}},\
  }\bibinfo {title} {On kernel target alignment},\ in\ \href
  {https://doi.org/10.1007/3-540-33486-6_8} {\emph {\bibinfo {booktitle}
  {Innovations in Machine Learning: Theory and Applications}}},\ \bibinfo
  {editor} {edited by\ \bibinfo {editor} {\bibfnamefont {D.~E.}\ \bibnamefont
  {Holmes}}\ and\ \bibinfo {editor} {\bibfnamefont {L.~C.}\ \bibnamefont
  {Jain}}}\ (\bibinfo  {publisher} {Springer Berlin Heidelberg},\ \bibinfo
  {address} {Berlin, Heidelberg},\ \bibinfo {year} {2006})\ pp.\ \bibinfo
  {pages} {205--256}\BibitemShut {NoStop}%
\bibitem [{\citenamefont {Hubregtsen}\ \emph {et~al.}(2022)\citenamefont
  {Hubregtsen}, \citenamefont {Wierichs}, \citenamefont {Gil-Fuster},
  \citenamefont {Derks}, \citenamefont {Faehrmann},\ and\ \citenamefont
  {Meyer}}]{Hubregtsen2022}%
  \BibitemOpen
  \bibfield  {author} {\bibinfo {author} {\bibfnamefont {T.}~\bibnamefont
  {Hubregtsen}}, \bibinfo {author} {\bibfnamefont {D.}~\bibnamefont
  {Wierichs}}, \bibinfo {author} {\bibfnamefont {E.}~\bibnamefont
  {Gil-Fuster}}, \bibinfo {author} {\bibfnamefont {P.-J. H.~S.}\ \bibnamefont
  {Derks}}, \bibinfo {author} {\bibfnamefont {P.~K.}\ \bibnamefont
  {Faehrmann}},\ and\ \bibinfo {author} {\bibfnamefont {J.~J.}\ \bibnamefont
  {Meyer}},\ }\bibfield  {title} {\bibinfo {title} {Training quantum embedding
  kernels on near-term quantum computers},\ }\href
  {https://doi.org/10.1103/PhysRevA.106.042431} {\bibfield  {journal} {\bibinfo
   {journal} {Phys. Rev. A}\ }\textbf {\bibinfo {volume} {106}},\ \bibinfo
  {pages} {042431} (\bibinfo {year} {2022})}\BibitemShut {NoStop}%
\bibitem [{\citenamefont {Wang}\ \emph {et~al.}(2021)\citenamefont {Wang},
  \citenamefont {Du}, \citenamefont {Luo},\ and\ \citenamefont
  {Tao}}]{Wang2021}%
  \BibitemOpen
  \bibfield  {author} {\bibinfo {author} {\bibfnamefont {X.}~\bibnamefont
  {Wang}}, \bibinfo {author} {\bibfnamefont {Y.}~\bibnamefont {Du}}, \bibinfo
  {author} {\bibfnamefont {Y.}~\bibnamefont {Luo}},\ and\ \bibinfo {author}
  {\bibfnamefont {D.}~\bibnamefont {Tao}},\ }\bibfield  {title} {\bibinfo
  {title} {Towards understanding the power of quantum kernels in the {NISQ}
  era},\ }\href {https://doi.org/10.22331/q-2021-08-30-531} {\bibfield
  {journal} {\bibinfo  {journal} {Quantum}\ }\textbf {\bibinfo {volume} {5}},\
  \bibinfo {pages} {531} (\bibinfo {year} {2021})}\BibitemShut {NoStop}%
\bibitem [{\citenamefont {Peruzzo}\ \emph {et~al.}(2014)\citenamefont
  {Peruzzo}, \citenamefont {McClean}, \citenamefont {Shadbolt}, \citenamefont
  {Yung}, \citenamefont {Zhou}, \citenamefont {Love}, \citenamefont
  {Aspuru-Guzik},\ and\ \citenamefont {O'Brien}}]{Peruzzo2014}%
  \BibitemOpen
  \bibfield  {author} {\bibinfo {author} {\bibfnamefont {A.}~\bibnamefont
  {Peruzzo}}, \bibinfo {author} {\bibfnamefont {J.}~\bibnamefont {McClean}},
  \bibinfo {author} {\bibfnamefont {P.}~\bibnamefont {Shadbolt}}, \bibinfo
  {author} {\bibfnamefont {M.-H.}\ \bibnamefont {Yung}}, \bibinfo {author}
  {\bibfnamefont {X.-Q.}\ \bibnamefont {Zhou}}, \bibinfo {author}
  {\bibfnamefont {P.~J.}\ \bibnamefont {Love}}, \bibinfo {author}
  {\bibfnamefont {A.}~\bibnamefont {Aspuru-Guzik}},\ and\ \bibinfo {author}
  {\bibfnamefont {J.~L.}\ \bibnamefont {O'Brien}},\ }\bibfield  {title}
  {\bibinfo {title} {A variational eigenvalue solver on a photonic quantum
  processor},\ }\href {https://doi.org/10.1038/ncomms5213} {\bibfield
  {journal} {\bibinfo  {journal} {Nature Communications}\ }\textbf {\bibinfo
  {volume} {5}},\ \bibinfo {pages} {4213} (\bibinfo {year} {2014})}\BibitemShut
  {NoStop}%
\bibitem [{qis(2023)}]{qiskitruntime}%
  \BibitemOpen
  \href@noop {} {\bibinfo {title} {{Getting started -- Qiskit Runtime IBM
  Client 0.14.0}}},\ \bibinfo {howpublished}
  {\url{https://qiskit.org/ecosystem/ibm-runtime/getting_started.html}}
  (\bibinfo {year} {2023}),\ \bibinfo {note} {accessed: 2023-11-14}\BibitemShut
  {NoStop}%
\end{thebibliography}
\end{document}